\documentclass[10pt]{article} 
\usepackage[preprint]{tmlr}

\usepackage{tablefootnote}
\usepackage{hyperref}
\usepackage{url}     
\usepackage[utf8]{inputenc} 
\usepackage{booktabs}       
\usepackage{microtype}      

\usepackage{chemformula}    
\usepackage[separate-uncertainty=true,exponent-product=\cdot]{siunitx}        
\DeclareSIUnit\angstrom{\text{\AA}} 
\usepackage{graphicx}
\usepackage{wrapfig}
\usepackage{xcolor}
\usepackage{mathtools}
\usepackage{todonotes}
\usepackage{xspace}
\usepackage{natbib}
\usepackage{makecell}
\usepackage{wrapfig}
\usepackage{tikz}
\usetikzlibrary[arrows.meta,bending]
\usetikzlibrary{positioning}
\tikzset{label/.style={fill=black,rounded corners,fill opacity=0.2,text opacity=1}}

\usepackage{caption} 
\usepackage[normalem]{ulem}

\newcommand{\model}[1]{\textsc{#1}\xspace}
\newcommand{\lorem}{\model{Lorem}}
\newcommand{\mace}{\model{Mace}}
\newcommand{\cace}{\model{Cace-Les}} 

\newcommand{\pet}{\model{Pet}}
\newcommand{\fgnn}{4G-NN\xspace}

\newcommand{\sn}{S\textsubscript{N}2\xspace}

\newcommand{\abr}[1]{\uppercase{#1}}

\newcommand{\mpnns}{\abr{mpnn}s\xspace}

\newcommand{\mlp}{\abr{MLIP}\xspace}
\newcommand{\mlps}{\abr{MLIP}s\xspace}

\newcommand{\gnns}{\abr{GNN}s\xspace}

\newcommand{\pes}{\abr{pes}\xspace}

\newcommand{\scare}[1]{\lq#1\rq}

\newcommand{\yes}{\checkmark}
\newcommand{\no}{$\times$}

\usepackage{amsmath,amsfonts,bm}

\def\eqref#1{equation~\ref{#1}}

\def\1{\bm{1}}

\def\vc{{\bm{c}}}

\def\vf{{\bm{f}}}

\def\vn{{\bm{n}}}

\def\vr{{\bm{r}}}

\def\vw{{\bm{w}}}

\def\evw{{w}}

\def\mA{{\bm{A}}}

\def\mK{{\bm{K}}}

\def\mM{{\bm{M}}}

\def\mP{{\bm{P}}}

\def\mR{{\bm{R}}}

\def\mW{{\bm{W}}}

\DeclareMathAlphabet{\mathsfit}{\encodingdefault}{\sfdefault}{m}{sl}
\SetMathAlphabet{\mathsfit}{bold}{\encodingdefault}{\sfdefault}{bx}{n}
\newcommand{\tens}[1]{\bm{\mathsfit{#1}}}
\def\tA{{\tens{A}}}

\def\tC{{\tens{C}}}

\def\tQ{{\tens{Q}}}

\def\tS{{\tens{S}}}

\def\tU{{\tens{U}}}
\def\tV{{\tens{V}}}

\def\tY{{\tens{Y}}}

\def\gE{{\mathcal{E}}}

\def\gG{{\mathcal{G}}}

\def\gV{{\mathcal{V}}}

\def\sX{{\mathbb{X}}}
\def\sY{{\mathbb{Y}}}

\def\emR{{R}}

\def\emW{{W}}

\newcommand{\etens}[1]{\mathsfit{#1}}

\def\etA{{\etens{A}}}

\def\etC{{\etens{C}}}

\def\etQ{{\etens{Q}}}

\def\etS{{\etens{S}}}

\def\etV{{\etens{V}}}

\def\etY{{\etens{Y}}}

\newcommand{\R}{\mathbb{R}}

\newcommand{\ourvector}[1]{{\bm{#1}}}
\newcommand{\ourmatrix}[1]{{\bm{#1}}}

\newcommand{\cutoff}{r_{\text{c}}}
\newcommand{\lmax}{l_{\text{max}}}
\newcommand{\lrlmax}{l_{\text{max, LR}}}

\newcommand{\graph}{\gG}
\newcommand{\nodes}{\gV}
\newcommand{\edges}{\gE}

\newcommand{\grad}{\nabla}

\let\R\undefined
\newcommand{\R}{\vr}

\newcommand{\F}{\vf}

\newcommand{\offset}{\vn}

\newcommand{\curlyset}[2]{\{\, #1\, |\, #2 \, \}}
\newcommand{\miniset}[1]{\{\, #1 \, \}}

\newcommand{\wholes}{\mathbb{Z}}
\newcommand{\reals}{\mathbb{R}}

\newcommand{\bigo}[1]{O(#1)}

\newcommand{\inlinefrac}[2]{#1/#2}

\newcommand{\scalar}[1]{\ourmatrix{#1}}
\newcommand{\scaf}{\mP}

\newcommand{\sphf}{\tS}

\newcommand{\func}[2]{\text{#1}(#2)}
\newcommand{\fcutoff}{f_{\text{cut}}}

\newcommand{\sog}{\text{SO}(3)}
\newcommand{\seg}{\text{SE}(3)}
\newcommand{\og}{\text{O}(3)}
\newcommand{\eg}{\text{E}(3)}

\usepackage[capitalise]{cleveref}       

\title{Learning Long-Range Representations with Equivariant Messages}

\author{\name Egor Rumiantsev$^*$, Marcel F. Langer$^{*\dagger}$, Tulga-Erdene Sodjargal,  Michele Ceriotti, \\ 
Philip Loche$^\dagger$ \\
\addr Laboratory of Computational Science and Modeling\\
École Polytechnique Fédérale de Lausanne\\
1015 Lausanne, Switzerland\\
$^*$ Contributed equally\\ 
$^\dagger$ Corresponding authors. Contacts: marcel.langer@epfl.ch, philip.loche@epfl.ch
}

\def\month{MM}  
\def\year{YYYY} 
\def\openreview{\url{https://openreview.net/forum?id=XXXX}} 

\newcommand{\CL}[1]{#1}
\newcommand{\CLs}[1]{}

\newcommand{\ML}[1]{#1}

\begin{document}

\maketitle

\begin{abstract}
Machine learning interatomic potentials trained on first-principles reference data are becoming valuable tools for computational physics, biology, and chemistry.
Equivariant message-passing neural networks, including transformers, achieve state-of-the-art accuracy but rely on cutoff-based graphs, limiting their ability to capture long-range effects such as electrostatics or dispersion, as well as electron delocalization.
While long-range correction schemes based on inverse power laws of interatomic distances have been proposed, they are unable to communicate higher-order geometric information and are thus limited in applicability.
To address this shortcoming, we propose the use of equivariant, rather than scalar, charges for long-range interactions, and design a graph neural network architecture, \lorem, around this long-range message passing mechanism.
We consider several datasets specifically designed to highlight non-local physical effects, and compare short-range message passing with different receptive fields to invariant and equivariant long-range message passing.
Even though most approaches work for careful dataset-specific choices of their model hyperparameters, \lorem{} \CL{works consistently without such changes, with excellent benchmark performance}.
\end{abstract}

\section{Introduction}

Machine learning interatomic potentials (\mlps) for atomistic simulations are trained on quantum-mechanical simulations to predict energies and forces of new atomic structures. Most \mlps assume locality: The energy of each atom depends only on neighbors within a cutoff radius. This leads to linear scaling with respect to the number of atoms but fails for systems with long-range interactions, such as electrostatics, dispersion, or electron delocalization. \citep{bp2007,kfgb2021,ts2009,gc2019,ucsgpstm2021,hlhc2023}
Several approaches aim to overcome this limitation.
Message-passing graph neural networks (\mpnns, \cite{gsrvd2017}) overcome locality by iteratively exchanging information between neighboring atoms, but are still constrained by the number of message-passing steps, graph connectivity, and reduced information flow with increasing number of iterations \citep{cw2020,ay2020,npfc2022}.

An alternative approach is to use physics-inspired corrections to the total energy, written as an inverse power law of interatomic distances, $1/r^p$ , with $p=1$ for charge–charge, $p=3$ for dipole–dipole, and $p=6$ for dispersion.
Some models predict partial atomic charges, either using explicit charge labels and equilibration schemes, or learning them implicitly from energies and forces to include electrostatic terms \citep{ucgssm2021,kfgb2021,fzklsnmlbbit2022,plsk2023,mka2025,c2025}.
Physical long-range interactions can also serve as building block: For instance, \cite{kggg2023} propose Ewald message passing, i.e., the use of electrostatic interactions for message passing, and \cite{gc2019} propose the long-distance equivariant (LODE) framework that uses inverse power-law interactions to compute equivariant features \ML{from the scalar potential around each atom}.

In this work, we combine the strengths of physical inverse power law $1/r^p$ interactions with the ability of equivariant message passing to communicate higher-order geometric information. Inverse power-law interactions are well-defined for periodic systems, have physically meaningful asymptotic behavior, and can be computed efficiently using established techniques from computational physics.
Extending the idea of Ewald message passing, we treat charges as equivariant objects and use inverse power-law potentials as a mechanism for long-range communication.
Based on this mechanism, we design \lorem, an \mlp architecture that combines short- and long-range message passing.
\lorem offers consistently high accuracy across a series of long-range benchmark tasks, outperforming other short- and long-range message passing models.

Our contributions are:
\begin{itemize}
  \item We introduce an equivariant, global, long-range message passing mechanism that can leverage efficient methods from computational physics for asymptotic $\bigo{N \log N}$ scaling in periodic systems, and potentially $\bigo{N}$ scaling in non-periodic ones,
  \item We design a novel \mlp architecture, \lorem, around this mechanism,
  \item We conduct a series of experiments to probe the limits of equivariant short-range message passing to model long-range physics.
\end{itemize}

\section{Background}
\label{sec:background}

\paragraph{Machine learning interatomic potentials}
\label{sub:mlp}


Under the \cite{bo1927} approximation, which decouples the nuclear and the electronic degrees of freedom, the atoms in a molecule or material move on a potential energy surface (\pes)
$E = E\left(\curlyset{(\R_i,Z_i)}{i=1...N}\right)$
where $\R_i$ and $Z_i$ are positions and atomic numbers for the $N$ atoms; in a periodic system, i.e., materials or liquids, the arrangement of $N$ atoms is repeated periodically in space, described by three cell vectors $\curlyset{\vc_a}{a=1,2,3}$ and corresponding integer offsets $\curlyset{n_a}{a=1,2,3}$ so that each position $\R_i$ is associated with the replicas $\curlyset{\R_i + \sum_{a=1}^3 n_a \vc_a}{\vn \in \wholes^3}$. The energy is then computed for the positions in the unit cell ($\vn = 0$) considering their interactions with the infinite \scare{crystal} system.
The forces, which drive the dynamics of the atoms, are defined as derivatives of the energy $\F_i = - \grad_{\R_i} E$.
Traditionally, this \pes has been approximated by physics-inspired analytical expressions, called force fields, that are parametrized manually or through global optimization.
In the past decades, in tandem with the increasing availability of large datasets of quantum mechanical reference data, \mlps have emerged as a less computationally efficient, but more accurate, data-driven alternative.

\paragraph{Atomistic graph neural networks}
Most \mlps can be seen as graph neural networks (\gnns, \cite{bhbszmtrsfgsbgdvanldhwkbvlp2018}) acting on a description of an arrangement of atoms in space as a geometric graph $\graph=(\edges,\nodes)$ with edges $\edges$ corresponding to interatomic relative-position vectors $\R_{ij}$ and nodes (or vertices) $\nodes$ corresponding to atoms. Edges connect nodes that lie within a cutoff radius $\cutoff$ of each other. In periodic systems, $\R_{ij}$ are constructed to respect periodic boundary conditions, i.e., if a replica lies closer than an original atom, $\R_{ij}$ points to the replica. If more than one replica lies within the cutoff, multiple edges with different labels are drawn between nodes.
By restricting the range of interactions to neighbors on this graph, linear scaling with the number of atoms $N$ (at constant density) can be achieved; efficient scaling with system size is required to make \mlps practical for large-scale simulations.
\mlps typically predict $E$ as a sum over atomic contributions, $E=\sum_{i=1}^N E_i$, predicted from node features.

\paragraph{Invariance and equivariance}
The potential energy $E$ is invariant under
permutations, i.e., reordering, of atomic positions,
as well as global translations and rotations of the coordinate system.
In other words, $E$ is invariant under actions of the (special) Euclidean symmetry group $\seg$ applied to all positions (including replicas).\footnote{In fact, the energy is invariant under global inversions as well, i.e., under the action of the full Euclidean group $\eg$, composed of translations and rotations+inversions, $\og$. To simplify notation in this manuscript, we focus on proper scalars and tensors, i.e., irreducible representations of $\sog$ -- all arguments and operations can be readily generalized to $\og$.}
\mlps must respect these symmetries, at least approximately.
Translation invariance is respected by construction in atomistic \gnns{} through the use of relative-position vectors as edge labels.
Permutation invariance is typically ensured through commutative aggregation functions, for instance sums.
Finally, rotation invariance can either be learned through data augmentation (\cite{pc2023,lpc2024}), or ensured by requiring that internal features remain aware of their geometric meaning, i.e., that they are \emph{equivariant} to rotations.
While in principle, \mlps could be constructed from invariant features only, equivariant internal features have been found to improve accuracy and data efficiency by allowing the model to access orientation information (\cite{tskylkr2018,bmsgmkmsk2022,bksoc2022}).
A thorough discussion of equivariance for \mlps can be found in other works \citep{s2021,um2024};
essentially, we consider the transformation of internal features of the models under rotations $g \in \sog$ applied equally to all geometric inputs (positions and cell vectors), leading to a joint rotation of all bulk positions.
Consider the \mlp up to some hidden layer as a function $f: \sX \rightarrow \sY$, where $ \sX = {\reals^{3 \times (N+3)}}$ for periodic systems and $\sX = {\reals^{3\times N}}$ for molecules.
Equivariance is defined by $f \circ g = g \circ f$ for all $g \in \sog$: rotations can be equivalently applied before or after $f$.
To ensure that this is the case, internal features must be constructed as direct sum of different irreducible representations of $\sog$, indexed by $l$, combined with a feature (channel) dimension $c$: $\sY = \left( \oplus_{l=0}^{\lmax} \reals^{2l+1} \right) \otimes \reals^c$.
Rotations act as linear transformations on these irreducible representations.
We write spherical features as tensor $\tS$ with the last three indices the representation order $l=0,...,\lmax$, the component $m=-l,...,l$, and the channel index $c$. Such tensors are therefore ragged: Different $l$ correspond to different numbers of components $m$. The $l=0$ components are called scalars and are invariant under rotations. Collections of purely scalar features are also denoted as matrices $\mP$. In \cref{sec:apx-modules}, we describe a number of operations that can be applied to spherical features without disrupting equivariance.

\paragraph{Long-range interactions}
The potential energy $E$ arises from the many-body Schrödinger equation, which involves only Coulomb interactions between electrons and nuclei without any range separation.
Efficient approximations, such as empirical force fields or \mlps, typically restrict interactions to local environments, motivated by the nearsightedness principle of electronic matter (\cite{pk2005}), which states that electronic properties are insensitive to distant perturbations.
In the long-range regime, interactions reduce to inverse power laws $1/r^p$ of the interatomic distance.
Since in nature no fixed nearsightedness length scale exists, \mlps must capture both local many-body quantum effects, possibly extending beyond the model's cutoff $\cutoff$, and formally infinite-range electrostatic interactions.
Additional complexity arises from charge distributions that depend on distant atoms and from electron wavefunctions that may delocalize over large distances; see \cref{sec:experiments}.

\paragraph{Ewald summation}
\label{sec:apx-ewald_summation}
The evaluation of inverse power-law potentials $1/r^p$ has been the object of much study in computational physics and chemistry. The general task is to compute the potential $V_i = V(\R_i)$ at a given atomic position $\R_i$, induced by (generalized) point charges\footnote{For $p=1$, i.e., electrostatics, it is common to speak of charges. For $p>1$, for example dispersion ($p=6$), \scare{coefficients} is more common.} $q_j$ placed at the position of other atoms $j$:
\begin{equation}
    V_i = \sum_{j=1}^N \sum_{\vn \in \wholes^3} \frac{q_j}{|\R_i - (\R_j + n_1 \vc_1 + n_2 \vc_2 + n_3 \vc_3 )|^p}  \,. \label{eq:E_and_V}
\end{equation}
Where $|\cdot|$ denotes the vector norm. In periodic systems, \cref{eq:E_and_V} implies an infinite sum over replicas, denoted by $\sum_{\vn \in \wholes^3}$ and defined by the cell vectors $\vc_a$; for non-periodic systems (molecules), this sum and the extra term in the denominator are omitted. For $p\leq3$, the periodic, infinite, sum converges conditionally, i.e., convergence or divergence depends on the summation order. \citet{e1921} summation was developed to tackle this problem for electrostatics ($p=1$) and later extended to other exponents; its basic concept is to split $1/r$ into a short-ranged part, which converges fast in real space, and a long-range part, which is smooth, and therefore converges well, in reciprocal space.
%
%
A naive implementation of Ewald summation scales $\bigo{N^2}$, which can be brought down to $\bigo{N^{3/2}}$ by choosing the cutoffs to be proportional to the size of the simulation cell. 
This is, however, undesirable for \mlps, which are typically constructed and trained for a fixed cutoff radius.
Particle–mesh Ewald (PME, P3M) algorithms reduce this to $\bigo{N \log N}$ by interpolating charges onto a grid and employing the fast Fourier transform for the reciprocal-space part \citep{dyp1993,he2021}.
While such methods are standard in force fields, implementations in popular machine learning frameworks, PyTorch \citep{pgmlbcklgadkydrtcsfbc2019} and JAX \citep{bfhjlmnpvwz2018}, have only become available recently \citep{lhhxrhlc2025}.
In non-periodic systems, the naive evaluation of \cref{eq:E_and_V} scales $\bigo{N^2}$. However, methods like the fast multipole expansion can reduce this cost to $\bigo{N}$ \citep{gr1987}. 
Another approach is multi-level summation \citep{hwpsss2015}, which scales $\bigo{N}$ for non-periodic and periodic systems and has recently been implemented in JAX \citep{bsucm2025pre}.
For systems with fewer than thousands of atoms, naive implementations---Ewald summation for periodic systems and direct summation for non-periodic systems---are typically faster than more complex, but better-scaling, methods; this is benchmarked, for instance, in \citet{lhhxrhlc2025}. Since all systems considered in this work are smaller than this threshold, we use naive implementations throughout.

\section{Related work}
\paragraph{Equivariant message passing}
Bond-order potentials first introduced the idea of repeatedly updating atomic environments to extend interactions beyond the cutoff radius $\cutoff$ \citep{t1988,b1990}, a principle now central to modern \mlps. In \mpnns \citep{gsrvd2017}, atoms exchange messages over $M$ steps, so features and energies depend on neighbors within $M \cdot \cutoff$. Letting ${}^k\mP_i$ denote the features at atom $i$ and message-passing step $k$:
\begin{align}
    {}^{k+1}\mM_{i} &= \sum_j {}^{k}m({}^k\mP_i, {}^k\mP_j, \R_{ij})  \label{eq:message} \\
    {}^{k+1}\mP_i &= {}^{k}u({}^k\mP_i, {}^{k+1}\mM_i) \, , \label{eq:update}
\end{align}
with learned message and update functions $m$ and $u$.
Early models used invariant updates \citep{sksctm2017,xg2018}, later extended to equivariant ones \citep{ggg2019,sug2021,bksoc2022,bmsgmkmsk2022,fumc2024}.
Equivariant \mpnns are restricted in the choice of operations within the network, as they must retain equivariance throughout (see \cref{sec:apx-modules}).
Recent universal \mlps trained on big datasets \citep{bbcwzc2025,wdfgsbagklmscdrsuz2025,mbkptfplc2025pre,rvsggdn2025} sometimes replace message passing with local self-attention. While effective at capturing semi-local interactions, \mpnns cannot model true long-range effects, since distant atoms without intermediates never interact, and many steps reduce expressivity \citep{cw2020,ay2020,npfc2022}.

\paragraph{Physics-based long-range models}
Many long-range models explicitly add physics-inspired terms to $E$ that capture interactions decaying more slowly with distance.
A common example of such a form is given by \cref{eq:E_and_V}: $E^{\mathrm{long}} = \sum_i q_i V_i$  where \( q_i \) are latent per-atom descriptors (e.g., partial charges or polarizabilities) predicted by the model and $V_i$ is computed as an inverse power law of interatomic distances.
%
Usually, these \( q_i \) are predicted directly as scalar functions of local atomic features \citep{ucgssm2021,shscrm2021,lhhxrhlc2025,kkzc2024,jlx2025,kfdksucmt2025, c2025}.
A modification of this approach is to include the \( q_i \) in an equilibration scheme, thus allowing these descriptors to capture information otherwise missed due to their initial dependence on local environments \citep{kfgb2021,plsk2023,mka2025}.
\citet{gc2019,hlhc2023} propose to use physics-inspired kernels to compute long-range features, mathematically equivalent to a multipole expansion, but find that higher-order features contain little additional information.

\paragraph{Other long-range models}
Some models avoid handcrafted corrections and instead learn long-range interactions directly. Ewald message passing augments GNNs with Fourier-space invariants \citep{kggg2023}, while fully connected approaches use all-to-all distances \citep{csmt2018} or global attention \citep{ucgssm2021}, though these lose efficiency or distance information. Linear-scaling attention with geometric embeddings \citep{fcmu2024} enables global orientation exchange but needs symmetrization and has not yet been adapted to periodic systems. Alternatives include virtual nodes for global aggregation \citep{cvgrnc2025} or message passing in spherical harmonics space \citep{fum2022}. Despite approximations, most methods still struggle to bridge periodic and non-periodic systems.
Some methods outside atomistic modeling also aim to capture long-range effects \citep{drgpwlb2022, bgrbd2025, mpxclm2025, zwm2025}, but require further adaptation to include geometric information or periodicity.

\section{Equivariant long-range message passing}
\label{sec:lrmp}



Ewald summation, as introduced in \cref{sec:apx-ewald_summation} and \cref{eq:E_and_V}, allows the efficient, and convergent in period systems, evaluation of a potential $V_i$ at each atomic position, based on coefficients $q_j$ associated with all other atoms $j$ and a power of the the inverse distances $\inlinefrac{1}{r_{ij}^p}$.
This computation can be seen as a physics-inspired form of message passing (\cref{eq:message,eq:update}), with radial filter $\inlinefrac{1}{r_{ij}^p}$, neighbor feature $q_i$, and resulting message $V_i$; this correspondence was pointed out by \citet{kggg2023} and used by \citet{gc2019} to compute general long-range features.

To bring it more in line with standard message passing, the operation can be carried out in parallel across an extra feature dimension, promoting $q_i$ to an array $\ourvector{q}_i$. This allows to communicate more information, but in this form, this information is restricted to geometric invariants.
We propose to promote $q_i$ to an \emph{equivariant} tensor instead: $\etQ_{i,l,m}$. This results in equivariant messages, or potentials:
\begin{equation}
  \etV_{i,l,m} = \sum_{j=1}^N \sum_{\offset\in\wholes^3} \frac{\etQ_{j,l,m}}{|\R_i - (\R_j + n_1 \vc_1 + n_2 \vc_2 + n_3 \vc_3 )|^p} \,, \label{eq:Vi_equi}
\end{equation}
carrying out the Ewald summation over each spherical order $l$ and component $m$ in parallel.

To see that the result is equivariant, recall (see \cref{sec:apx-modules}) that adding two equivariant objects yields another equivariant object, and that multiplication with a prefactor, provided that the factor is shared across all entries for a given $l$, also retains equivariance. \CL{The prefactor $\inlinefrac{1}{r_{ij}^p}$ depends on the pair $(i,j)$ and therefore varies across pairs; however, for any given pair, it is a single scalar that multiplies all $m$ components equally within each $l$. Since the action of rotations on spherical features is a linear map along the $m$ index, this scalar multiplication commutes with it.} It is then easy to see that multiplying $\tQ_i$ with a factor $\inlinefrac{1}{r_{ij}^p}$ yields an equivariant quantity, and that the sum in \cref{eq:Vi_equi} also yields an equivariant, since all summands are equivariant objects. We argue and numerically confirm in \cref{sec:ap_equiv} that a compensating background charge correction preserves equivariance.

Overall, this approach allows the global aggregation of equivariant messages in a way that is amenable to efficient implementation, scaling $\bigo{N \log N}$ in periodic systems \ML{(see \cref{sec:apx-benchmark})} and---at least in principle, see \cref{sec:apx-ewald_summation}---$\bigo{N}$ in finite systems. Even more advanced computational methods can bring $\bigo{N}$ scaling to both. The method also naturally accommodates the long-range effects in physical systems, which decay asymptotically following a power law.

\section{\lorem}

\begin{figure}[h]
\begin{center}
\includegraphics{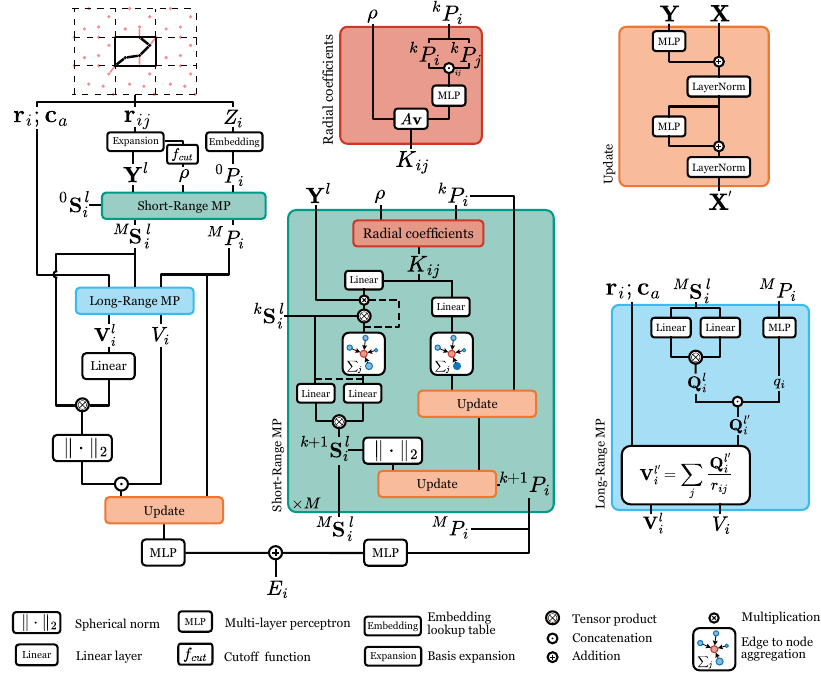}
\end{center}
\caption{Sketch of the \lorem architecture.}
\label{fig:lorem}
\end{figure}

\lorem follows the blueprint of equivariant \mpnns, making a few modifications: Scalar and spherical features are handled separately, following \citet{fum2022,fumc2024}, and mixed using a variant of the power spectrum \citep{bkc2013} and a residual update block \citep{hzrs2015}.

\paragraph{Overview} \lorem, illustrated in \cref{fig:lorem} maps a point cloud of $N$ atomic positions $\miniset{\R_i}$, potentially contained in a periodic unit cell $\vc_1,\vc_2,\vc_3$, and labeled with chemical species $\miniset{Z_i}$, to atomic energies $\miniset{E_i}$ which, when summed, yield the total potential energy $E$. Internally, the point cloud is processed as a graph, with edges between nodes, the atoms, defined by a cutoff radius $\cutoff$. Accordingly, edges correspond to vectors $\R_{ij}=\R_j-\R_i$ connecting atoms $i$ and $j$. Node features are updated through either {\bf short-range message passing} or {\bf long-range message passing}. Spherical information is used to update scalar features by computing their {\bf spherical norm} and passing it through an {\bf update} block. Updates of scalar features are followed by a residual prediction of atomic energy contributions.
We discuss the main components of \lorem below; specialized equivariant operations are described in more detail in \cref{sec:apx-modules}.


\paragraph{Short-range message passing}

Initial scalar features ${}^0\scaf_i$ are a learned embedding of chemical species.
At each short-range message passing step $k$, edge features $\scalar{K}_{ij}$ are obtained from distances $r_{ij}$ and scalar features ${}^k\scaf_i$ and ${}^k\scaf_j$ through a {\bf radial expansion}.
These edge features are then linearly transformed twice, with different learned weight matrices: Once to yield scalar messages that are aggregated into scalar updates, and again to yield pre-factors for the spherical harmonics $\tY_{ij}$ ({\bf angular expansion}), which are combined with neighboring spherical features $\tS_j$ in a tensor product.\footnote{In the initial message passing step, no spherical node features are available, and therefore the tensor product with $\tS_j$ is omitted. Initial spherical node features are obtained via a self tensor product (dotted lines) rather than a tensor product of the updates with the previous features.}
The results are aggregated into a spherical message, which updates the spherical features via another {\bf tensor product}, resulting in ${}^{k+1}\sphf_i$.
The spherical norm of the updated spherical node features is then used to further update the scalar features, finally yielding  ${}^{k+1}\scaf_i$.
This process is repeated $M$ times (the total number of short-range message passing steps).

\paragraph{Long-range message passing} We use the method discussed in \cref{sec:lrmp} to communicate equivariant information beyond the effective interaction cutoff of short-range message passing. To minimize computational cost, which is proportional to the number of channels over which Ewald summation is carried out, node features are transformed into low-dimensional charges: The scalar features ${}^M\scaf_i$ are transformed into a single per-atom charge $q_i$, and spherical features ${}^M\sphf_i$ are likewise, using a linear transformation followed by a self tensor product, transformed into $\etQ_{i,l,m}$ with a low $\lrlmax$ and a singular feature dimension. Unless otherwise noted, we use $\lrlmax=2$. The scalar charge is concatenated with the $l=0$ spherical charge, which yields a total of $10$ charge channels for $\lrlmax=2$.\footnote{Two with $l{=}0$: one from the scalar features and one from the scalar component of the spherical features.}
After Ewald summation, which is carried out in parallel across $l$ and $m$, the potentials are split back into the purely scalar $V_i$ and the spherical $\tV_i$. The spherical potentials are combined with spherical node features through a tensor product; the spherical norm of the result is concatenated with the scalar potential to update scalar representations.
We find that using only $p{=}1$, i.e., the Coulomb interaction, rather than a set of different exponents, is sufficient in practice. Experiments for $\lrlmax=0,1,2$ can be found in \cref{sec:apx-ablations}.


\paragraph{Update block} Updates $\ourvector{Y}$ to scalar features $\ourvector{X}$ use an update block consisting of multi-layer perceptrons (MLP) and layer normalization (LayerNorm) \citep{bkh2016}, following a residual structure \citep{hzrs2015}:
\begin{align}
  \ourvector{X} &\leftarrow \ourvector{X} + \func{MLP}{\ourvector{Y}} \nonumber\\
  \ourvector{X} &\leftarrow \func{LayerNorm}{\ourvector{X}} \nonumber\\
  \ourvector{X} &\leftarrow \ourvector{X} + \func{MLP}{\ourvector{X}} \nonumber\\
  \ourvector{X} &\leftarrow \func{LayerNorm}{\ourvector{X}} \nonumber \, .
\end{align}

\paragraph{Radial expansion} \lorem processes information about the distance between two atoms $r_{ij}$ through learned coefficients for linear transformations of an initial radial basis expansion of $r_{ij}$. Distances $r_{ij}$ are first expanded in Bernstein polynomials multiplied with $\fcutoff$, the cosine cutoff function (extending from $0$ to $\cutoff$), yielding initial radial features $\rho_{ij,c}$. These features are multiplied with a weight matrix $\mA$, which is in turn obtained through a MLP (and subsequent reshaping operation) applied to the concatenated scalar atom features $\scaf_i$ and $\scaf_j$. This allows the model to learn a radial basis based on the features of both atoms $i$ and $j$. The resulting edge features are called $\mK_{ij}$.

\paragraph{Angular expansion} The vectors $\R_{ij}$ are expanded in spherical harmonics $\tY_{ij}=\tY(\R_i)$, which are polynomials of vector components that produce outputs in different irreducible representations of $\sog$ \citep{um2024}. During message passing, they are multiplied with per-$l$ coefficients: $\etY'_{ij,l,m,c} = \etA_{ij,l,c} \etY_{ij,l,m}$, where $\tA_{ij}$ are obtained as linear transformation of $\mK_{ij}$, followed by reshaping operation. Since the factors are shared across $m$, this preserves equivariance.

\label{sec:model}

\section{Experiments}
\label{sec:experiments}

We perform experiments on a number of existing benchmark tasks designed to probe the ability of \mlps to model long-range interactions, comparing \lorem to short and long-range \mlps{}.
The experiments are divided into two parts: In \cref{sub:default}, we compare \lorem with other models using standardized settings. We find that \lorem{} performs well, but also observe that most tasks can be solved by models that do not consider long-range interactions at all -- the effective interaction range of typical message passing models is sufficient. However, predictions break down beyond this interaction range.
In \cref{sub:limit}, we probe this limit of short-range message passing. We find that while careful consideration of the number of message-passing steps and the cutoff radius is required to resolve long-range interactions with short-range models, \lorem can solve these benchmarks without adaptation. \CL{Additional experiments can be found in the appendix: Runtime benchmarks comparing Ewald and particle-mesh Ewald implementations of \lorem, demonstrating near-linear scaling up to \qty{30}{k} atoms in \cref{sec:apx-benchmark}, ablations of $\lrlmax$ and the long-range block in \cref{sec:apx-ablations}, }\ML{ as well as competitive performance on the larger ADAPT dataset \citep{dxzhrjk2025pre} in \cref{sec:apx-adapt}.}


\paragraph{Models}
We compare \lorem with a number of purely short-ranged \mlps, \mace and \pet, as well as the recently introduced \cace model that combines short-range message passing with a scalar long-range part, and \fgnn, which includes a physics-based long-range energy contribution and charge equilibration. \CL{Additionally we include metrics for SpookyNet \citep{ucgssm2021}, which predicts scalar partial charges and nuclear dipoles for long-range corrections, together with global attention.} Full details on model descriptions and training can be found in \cref{sec:apx-models}; \CL{approximate parameter counts are: \lorem ${\sim}\qty{1}{M}$, \cace  ${\sim}\qty{70}{k}$, \mace  ${\sim}\qty{800}{k}$, \pet ${\sim}\qty{1}{M}$, \fgnn ${\sim}\qty{5}{k}$ (estimated), \textsc{SpookyNet} ${\sim}\qty{3}{M}$ (from \citep{bmc2023}).}

\begin{table}
    \caption{Root mean squared errors for energy $E$ and forces $\F$ for datasets and models used in \cref{sub:default}. Where available, a held-out test set was used; otherwise, the validation set was used instead and indicated in the table. The lowest error is indicated in bold, the second lowest is underlined. The second line after each model name indicates the number of short-range (SR) message-passing steps and whether some form of long-range (LR) interactions are included in the model.\tablefootnote{\fgnn training requires DFT‑computed charges, which are not available for the biodimers and cumulene datasets.}
    An MAE version of this table is provided in \cref{sec:apx-mae-rmse}.}
    \centering
    \begin{tabular}{r | r r r r r r}
\toprule
                 Dataset  &  \makecell{\textsc{Lorem} \\{ $1\times$SR$+$LR}}  &  \makecell{\textsc{Cace-Les} \\{ $1\times$SR$+$LR}}  &  \makecell{\textsc{Mace} \\{ $2\times$SR}}  &  \makecell{\textsc{Pet} \\{ $2\times$ SR}}  &  \makecell{4G-NN \\{ $1\times$SR$+$LR}}  &  \makecell{\textsc{SpookyNet} \\{ $6\times$SR$+$LR}} \\ 
\midrule
\ch{MgO} surface\hspace{3mm} $E$ (meV/at)  &      \textbf{\num{0.064}}  &   \underline{\num{0.071}}  &               \num{0.376}  &               \num{0.210}  &               \num{0.219}  &               \num{0.107} \\ 
{\small (Validation)}\hspace{3mm} $\F$ (meV/Å)  &      \textbf{\num{4.076}}  &               \num{7.913}  &               \num{5.971}  &               \num{6.261}  &              \num{66.000}  &   \underline{\num{5.337}} \\ 
\midrule
\ch{NaCl} cluster\hspace{3mm} $E$ (meV/at)  &      \textbf{\num{0.112}}  &               \num{0.210}  &               \num{1.681}  &               \num{1.517}  &               \num{0.481}  &   \underline{\num{0.135}} \\ 
{\small (Validation)}\hspace{3mm} $\F$ (meV/Å)  &   \underline{\num{1.155}}  &               \num{9.784}  &              \num{40.219}  &              \num{42.438}  &              \num{32.780}  &      \textbf{\num{1.052}} \\ 
\midrule
Biodimers\hspace{3mm} $E$ (meV/at)  &      \textbf{\num{0.222}}  &   \underline{\num{2.259}}  &               \num{7.793}  &               \num{6.758}  &                        --  &                        -- \\ 
\hspace{3mm} $\F$ (meV/Å)  &      \textbf{\num{1.646}}  &   \underline{\num{3.163}}  &              \num{16.150}  &              \num{16.470}  &                        --  &                        -- \\ 
\midrule
Cumulene\hspace{3mm} $E$ (meV/at)  &   \underline{\num{3.309}}  &              \num{17.803}  &              \num{12.592}  &      \textbf{\num{3.205}}  &                        --  &                        -- \\ 
\hspace{3mm} $\F$ (meV/Å)  &  \underline{\num{50.084}}  &             \num{147.616}  &             \num{104.318}  &     \textbf{\num{46.905}}  &                        --  &                        -- \\ 
\bottomrule
\end{tabular}

    \label{tab:metrics}
\end{table}

The datasets and associated benchmark tasks used in our experiments are briefly introduced below; additional details are given in \cref{sec:ap_datasets}. An overview of validation or test set (where available) metrics for all datasets is given in \cref{tab:metrics}. \lorem performs very well across datasets, and is competitive with, or more accurate than, other models.

\paragraph{\ch{MgO} surface} This benchmark task is the first in a series designed by \cite{kfgb2021} to highlight the need for long-range information in \mlps. Illustrated in \cref{fig:aumgo+nacl}A, it consists of a magnesium oxide (\ch{MgO}) surface on which a gold (\ch{Au_2}) dimer is placed. Depending on the presence of an aluminum (\ch{Al}) dopant deep inside the surface, the lowest-energy position of the gold dimer is either in the \scare{wetting} (flat), or \scare{non-wetting} (standing up), position.
The benchmark consists of two parts: Correctly identifying the ordering between upright and flat geometries in the doped and undoped case, and reproducing the energy-distance curve for the non-wetting geometry, in particular the local minimum corresponding to the equilibrium distance.

\paragraph{NaCl cluster} Similar to the presence of a dopant modifying the potential-energy surface for the gold dimer in the \ch{MgO} surface task, this benchmark by \cite{kfgb2021} relies on the presence or absence of a sodium atom at one end of a charged sodium chloride (\ch{NaCl}) cluster changing the behavior of a sodium atom at the opposite end: Since one atom is removed while the charge remains constant, the charge must redistribute over the remaining atoms.
The benchmark task consists of reproducing the location of the local minimum and the energy profile when moving the sodium atom farthest from the removed one, indicated in \cref{fig:aumgo+nacl}C.

\paragraph{Cumulene} The cumulene benchmark task was proposed by \cite{ucsgpstm2021} as an example of a long-range problem that is due to a non-local effect of the electronic structure of a molecule. This molecule consists of a chain of nine carbon atoms, with a pair of hydrogen atoms, the rotors, at the opposite ends. The orientation of the rotors determines the shape of the atomic orbitals for each carbon, which propagates along the chain to the other end; in the absence of bending or stretching, the energy is therefore fully determined by the relative orientation of the rotors. The task is to recover the energy profile of this rotation for an idealized fully extended geometry, illustrated in \cref{fig:cumulene}.

\paragraph{Biodimers} This benchmark dataset by \cite{hlhc2023} consists of pairs of relaxed organic molecules placed at distances of \qtyrange{4}{15}{\angstrom} from each other, illustrated in \cref{fig:bio_dimers}A. Depending on the chemical nature of the molecules, interactions between the molecules take different asymptotic power-law forms, ranging from charge-charge $1/r$ interactions to apolar-apolar $1/r^6$ interactions. Here, we simply compare energy and force prediction errors on test sets stratified by the dominating power-law interaction.

\paragraph{\sn reactions} Finally, the \sn reactions benchmark was introduced by \cite{fcmu2024} to probe the ability of \mlps to model the long-range interactions required to mediate gas-phase chemical reactions. It consists of the nucleophilic substitution of methyl halides by another halide ion: $\ch{X^-} + \ch{H_3C-Y} \rightarrow \ch{X-CH_3} + \ch{Y^-}$ where \ch{X,Y} = \ch{F}, \ch{Cl}, \ch{Br}, or \ch{I}. The benchmark task in this case is predicting the energy along the reaction coordinate, i.e., correctly modeling the potential energy profile as the two reactants approach one another, react, and separate again.


\subsection{Standardized settings}
\label{sub:default}

We compare \lorem to other models using standardized settings: Two message-passing steps for short-range models, one message-passing step for long-range models, and $\cutoff=\qty{5}{\angstrom}$ for most models (see \cref{sec:apx-models}).

\begin{figure}
  \centering
    \begin{tikzpicture}[
        x = 1cm,
        y = 1cm,
        ]
        \node (figure) at (4,0) {\includegraphics[scale=0.65]{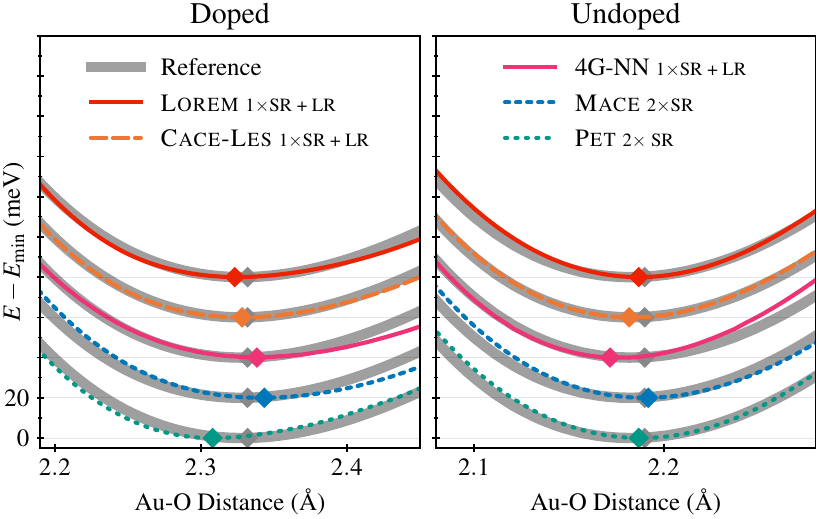}};
        \node (render) at (-3.2,0.1) {\includegraphics[scale=0.18]{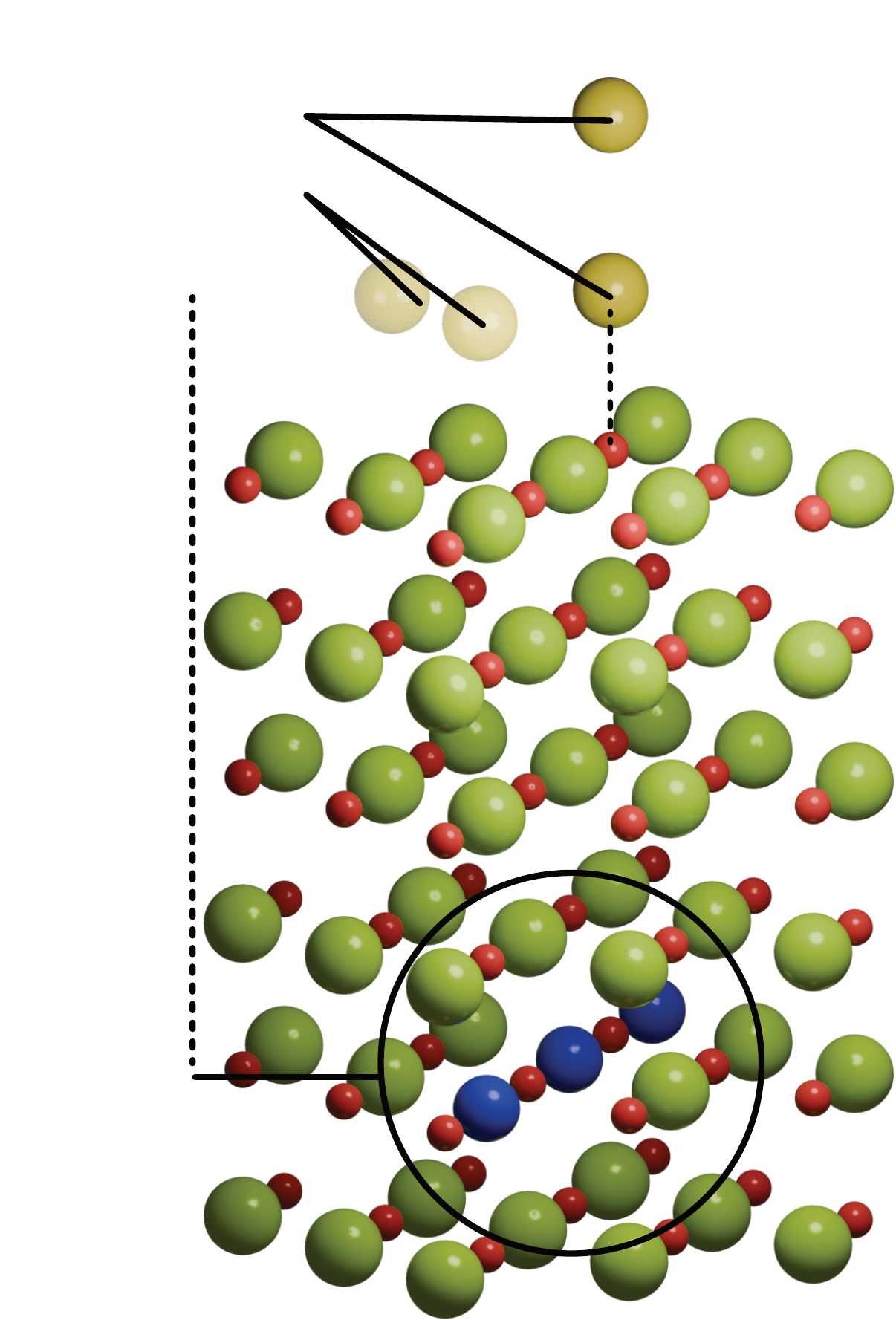}};
        \node (l) at (-4.80, 0.20) {$\sim\hspace{-0.5mm}\qty{12}{\angstrom}$};
        \node (d) at (-2.35, 1.4) {$d$};
        \node (w) at (-4.70, 2.35) {non-wetting};
        \node (nw) at (-4.40, 2.00) {wetting};
        \node (dp) at (-4.80, -1.6) {dopant};
        \node[label] () at (-5, 2.9) {$\mathbf{A}$};
        \node[label] () at (-0.5, 2.9) {$\mathbf{B}$};
        \node (figure) at (4,-5.25) {\includegraphics[scale=0.65]{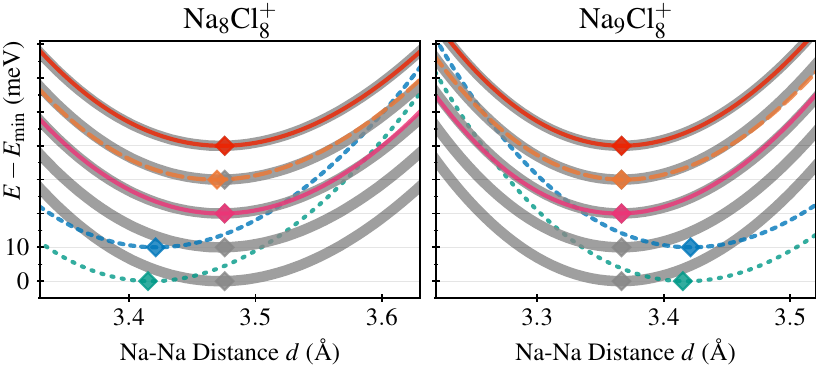}};
        \node (render) at (-3.2,0.1-5.25) {\includegraphics[scale=0.18]{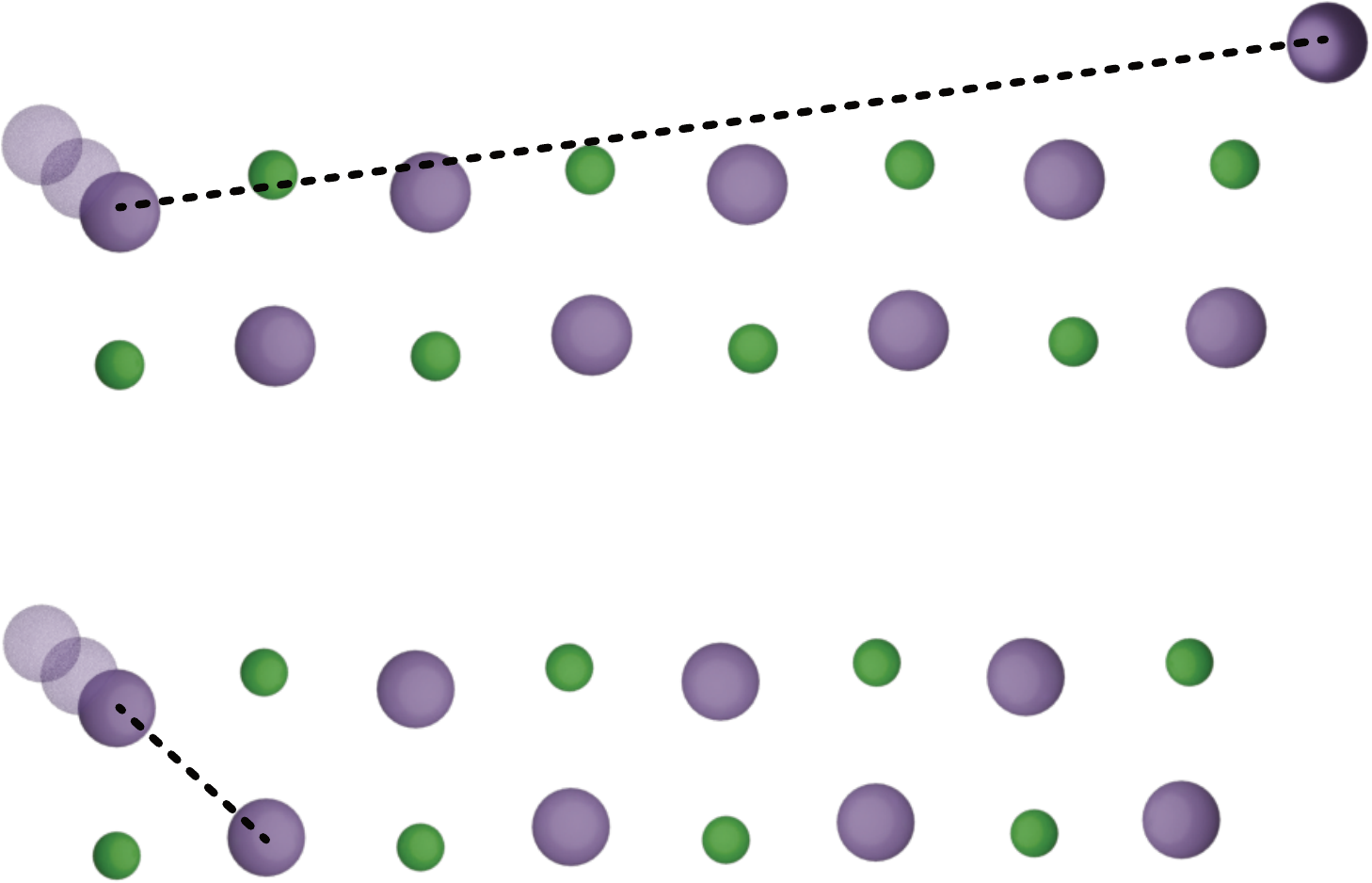}};
        \node (l) at (-3.15, 1.50-5.25) {$\sim\hspace{-0.5mm}\qty{21}{\angstrom}$};
        \node (d) at (-4.65, -1.50-5.25) {$d$};
        \node[label] () at (-5, 2.2-5.45) {$\mathbf{C}$};
        \node[label] () at (-0.5, 2.2-5.45) {$\mathbf{D}$};
    \end{tikzpicture}
    \caption{
    ($\mathbf{A}$) \ch{Au_2} dimer on \ch{MgO} surface, showing both wetting and non-wetting geometries, as well as the \ch{Al} dopant. ($\mathbf{B}$) Energy over distance $d$ for the non-wetting geometry for the doped and undoped surface. The minima are indicated with a diamond symbol; the reference energy curve is drawn in grey. Offsets are added to distinguish the curves and the value at the minimum is subtracted.
    ($\mathbf{C}$) \ch{Na_9Cl_8^+} (top) and \ch{Na_8Cl_8^+} (bottom) cluster, the moving atom is marked with transparent copies of itself, and the distance of interest is labeled with $d$.
    ($\mathbf{D}$) Energy over distance for both clusters.
    }
    \label{fig:aumgo+nacl}
\end{figure}

\paragraph{\ch{MgO} surface} The results for this experiment can be seen in \cref{fig:aumgo+nacl}B: Despite being designed to require long-range interactions, this benchmark task can be solved by both purely short-range and long-range message-passing models. 
In this case, the success of short-range message passing is due to the small size of this benchmark system: With an effective cutoff radius above $\qty{6}{\angstrom}$, centrally located atoms can \scare{see} the full system and hence, \mace and \pet can solve this task.
All models also resolve the orientation preference for the \ch{Au_2} dimer between the doped and undoped surfaces.

\paragraph{\ch{NaCl} cluster}
The results of this experiment, seen in \cref{fig:aumgo+nacl}D, are drastically different from the previous one: Here, only models with a long-range component are able to resolve the difference between \ch{Na_9Cl_8^+} and \ch{Na_8Cl_8^+}. All such models show excellent agreement with reference values. The failure of short-range message-passing is due to the larger system size compared to the \ch{MgO} surface: Here, effective cutoff radii exceeding \qty{10.5}{\angstrom} are required to solve the benchmark task.

\begin{figure}
  \centering
    \begin{tikzpicture}[
        x = 1cm,
        y = 1cm,
        C/.style = {circle, fill=gray, minimum size=8pt, inner sep=0pt, outer sep=2pt, draw=black, thick},
        H/.style = {circle, fill=red, minimum size=6pt, inner sep=0pt, outer sep=2pt, draw=black, thick},
        ]
        \node (figure) at (-3.2,0) {\includegraphics[scale=0.65]{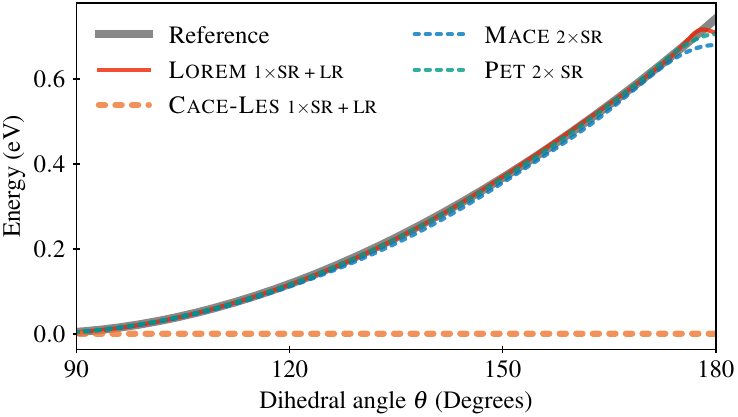}};
        \node (render) at (-4.5,0.05) {\includegraphics[scale=0.16, angle=-5]{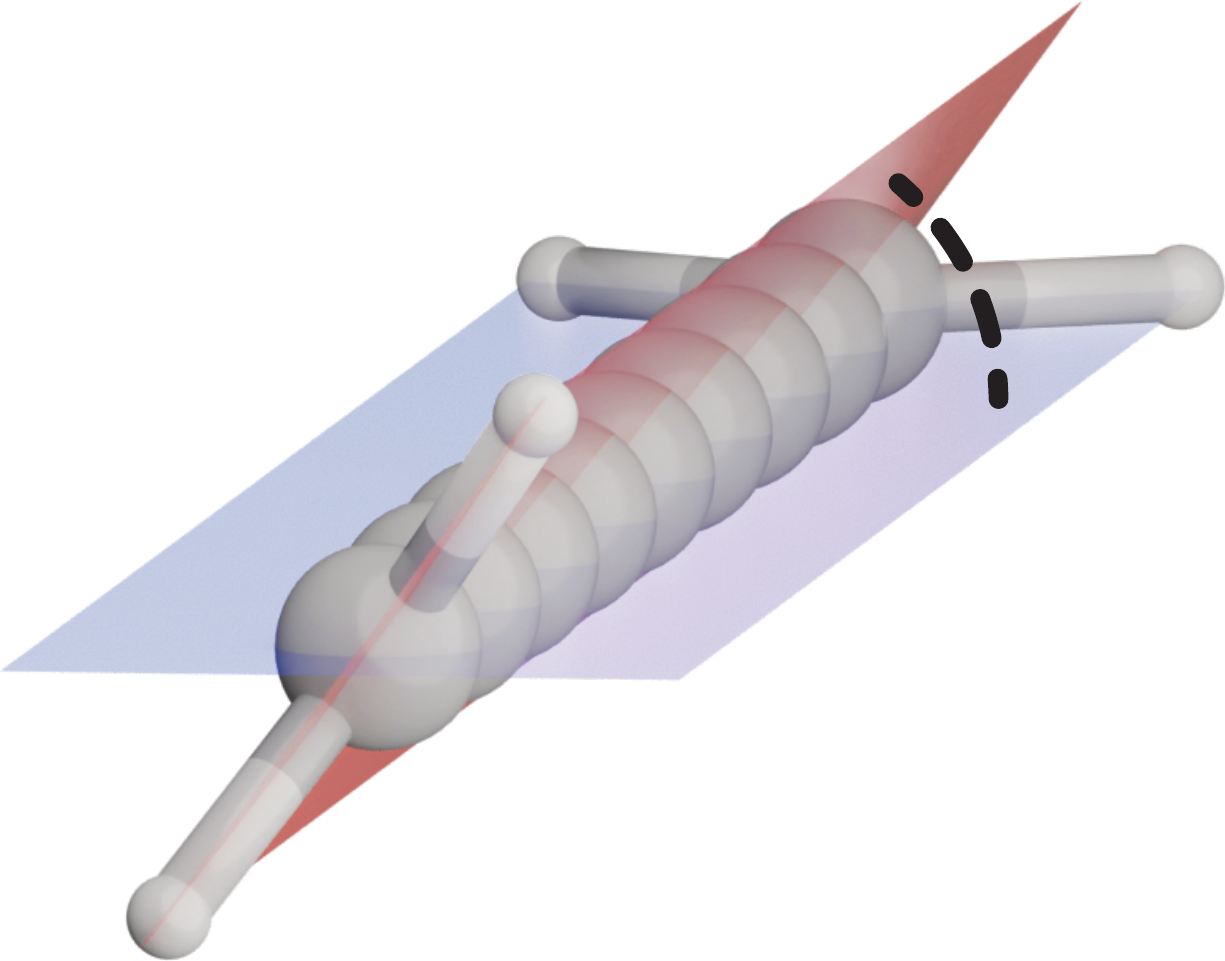} };
        \node (c) at (-3.3, 0.75) {$\theta$};

        \node[label] () at (-12.5, 2.8) {$\mathbf{A}$};
        \node[label] () at (-6.7, 2.8) {$\mathbf{B}$};

        \begin{scope}[shift={(-11,4)}, rotate=35]
          \node[H] (H0) at (-4-0.2,-0.55-3.5) {};
          \node[H] (H1) at (-4-0.2,+0.55-3.5) {};
          \node[C] (C0) at (-4+0,0-3.5) {};
          \node[C] (C1) at (-4+0.75,0-3.5) {};
          \node[C] (C2) at (-4+1.5,0-3.5) {};
          \node[C] (C3) at (-4+2.25,0-3.5) {};
          \node[C] (C4) at (-4+3.0,0-3.5) {};
          \node[C] (C5) at (-4+3.75,0-3.5) {};
          \node[C] (C6) at (-4+4.5,0-3.5) {};
          \node[C] (C7) at (-4+5.25,0-3.5) {};
          \node[C] (C8) at (-4+6.0,0-3.5) {};
          \node[H] (H2) at (-4+6.0+0.2,-0.55-3.5) {};
          \node[H] (H3) at (-4+6.0+0.2,+0.55-3.5) {};
        \end{scope}

        \draw [dotted,thick] (H1) to node [midway, xshift=-8, yshift=-8]{\qty{1.1}{\angstrom}} (C0) ;
        \draw [dotted,thick] (C2) to node [midway, xshift=-10, yshift=10]{\qty{1.3}{\angstrom}} (C3) ;
        \draw [dotted,thick] (H0) to node [midway, xshift=15, yshift=0]{\qty{2.1}{\angstrom}} (C1) ;
        \draw [dotted,thick] (C6) to node [midway, xshift=-10, yshift=10]{\qty{3.3}{\angstrom}} (H3) ;
        \draw [dotted,thick] (C4) to node [midway, xshift=0, yshift=-10]{\qty{5.8}{\angstrom}} (H2) ;

        \draw [solid,thick] (H0) -- (C0) ;
        \draw [solid,thick] (C0) -- (C1) ;
        \draw [solid,thick] (C1) -- (C2) ;
        \draw [solid,thick] (C3) -- (C4) ;
        \draw [solid,thick] (C4) -- (C5) ;
        \draw [solid,thick] (C5) -- (C6) ;
        \draw [solid,thick] (C6) -- (C7) ;
        \draw [solid,thick] (C7) -- (C8) ;
        \draw [solid,thick] (C8) -- (H2) ;
        \draw [solid,thick] (C8) -- (H3) ;

    \end{tikzpicture}
    \caption{
    ($\mathbf{A}$)~Illustration of cumulene laid flat, indicating relevant distances between atoms.
    ($\mathbf{B}$)~Energy profile over a \qty{90}{\degree} rotation of one rotor. The minimum value of each curve is subtracted before plotting. The inset shows a 3D representation of cumulene, defining the dihedral angle $\theta$.
    }
    \label{fig:cumulene}
\end{figure}

\paragraph{Cumulene} Resolving the cumulene energy profile, seen in \cref{fig:cumulene}, requires simultaneous knowledge of the orientation of both rotors at opposite ends of the molecule. This can be achieved in two ways: Through equivariant short-range message passing (\mace, \pet), provided that the chain is not too long (see \cref{sub:limit}), or through \emph{equivariant} long-range message passing (\lorem). For this reason,  \cace cannot solve this benchmark: Scalar charges are not sufficiently expressive to communicate relative orientation.
All equivariant models that are able to access this information can solve this benchmark, achieving good agreement with the reference data. 
\pet resolves the dihedral angle as well, but requires long training and an adaptation in the number of transformer layers to succeed at this benchmark (see \cref{sec:apx-models}). 
We note that the sharp cusp at \qty{180}{\degree} is an artifact of the underlying reference method; it is a desirable behavior of \mlps to smoothen it out.

\begin{figure}
  \centering
    \begin{tikzpicture}[
        x = 1cm,
        y = 1cm,
        ]
        \node (figure) at (6.0,0) {\includegraphics[scale=0.65]{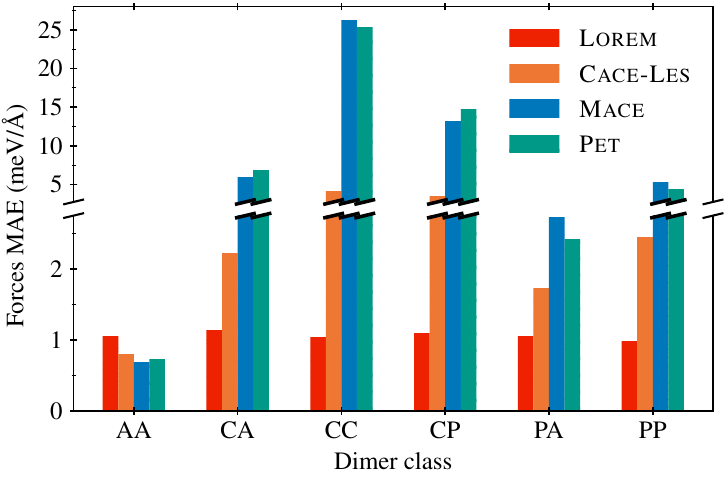} };
        \node (render) at (-1,0.1) {\includegraphics[scale=0.18,angle=-30]{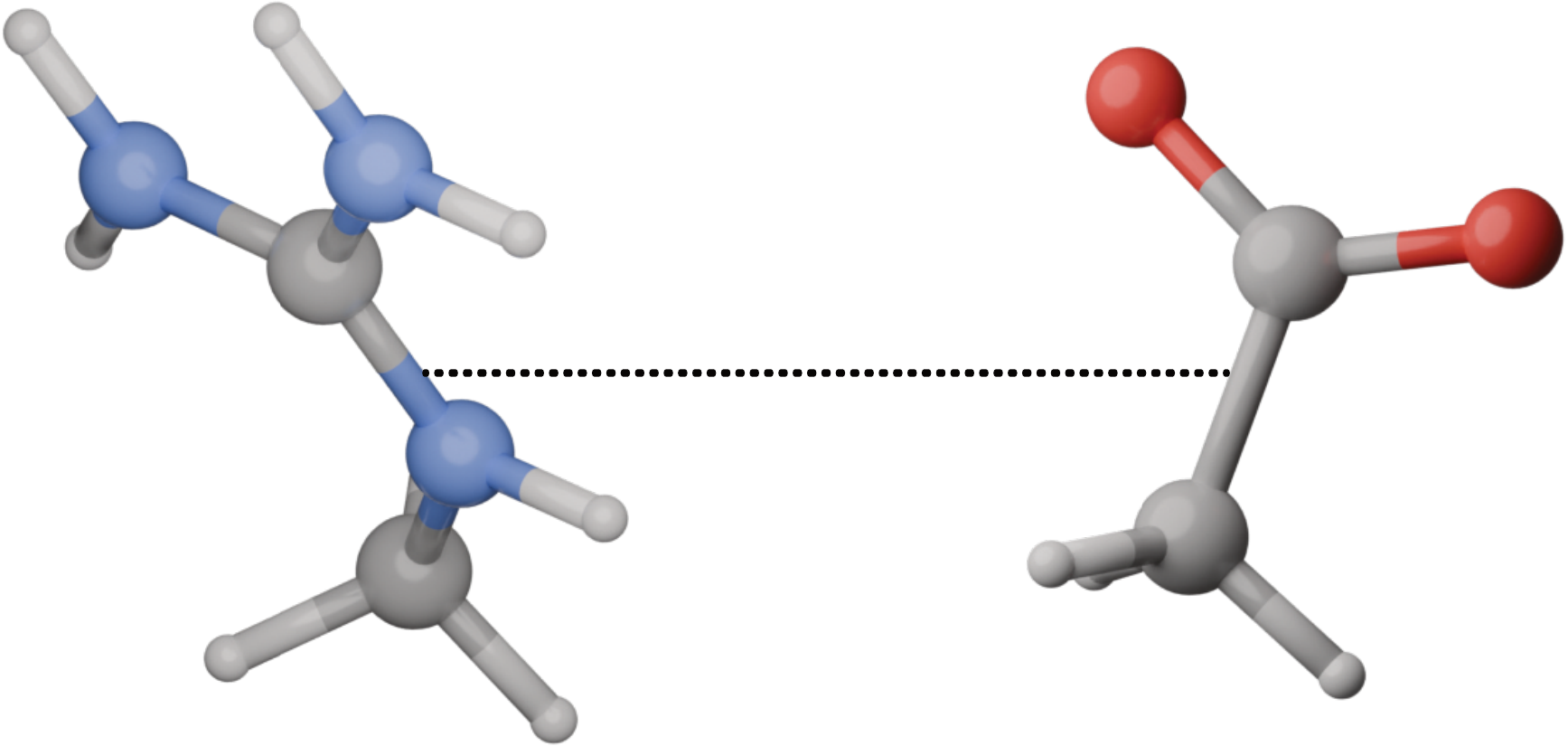} };
        \node[rotate=-32] (l) at (-0.9, 0.4) {$\sim\hspace{-0.5mm}\qty{4}{\angstrom} -\qty{15}{\angstrom}$};
        \node[label] () at (-3.5,3.0) {$\mathbf{A}$} ;
        \node[label] () at (2.2,3.0) {$\mathbf{B}$} ;
    \end{tikzpicture}
    \caption{
    ($\mathbf{A}$) Charge-charge pair from the biodimers dataset.
    ($\mathbf{B}$) Mean absolute error on forces for different models on the different dimer classes: Apolar-apolar (AA), charge-apolar (CA), charge-charge (CC), charge-polar (CP), polar-apolar (PA), polar-polar (PP). Note that the vertical axis has been split at \qty{2.75}{meV\per\angstrom}.
    }
    \label{fig:bio_dimers}
\end{figure}

\paragraph{Biodimers}
Since the pairs of molecules in the biodimers benchmark are placed at separations up to \qty{15}{\angstrom}, much beyond the cutoff used for graph construction, high accuracy requires a long-range component.
This is confirmed by the results of this experiment, seen in \cref{fig:bio_dimers}: The long range \lorem and \cace models yield lower error than \mace and \pet. Predictive error varies between dimer classes, i.e., the expected type of inverse power-law interaction, for all models, with the exception for \lorem, which yields consistent, and in all but one class the highest, accuracy. \ML{It is important to }\CL{note that \lorem uses $p=1$ (Coulomb) for long-range message passing while different exponents describe the \emph{underlying physics} of the dimer classes: The nonlinearity after the long-range message passing block may allow the model to correct this mismatch between exponents;} \ML{see also the supplement of \citep{hlhc2023}.} 

\subsection{Limits of short-range message passing}
\label{sub:limit}

In the previous experiments, we observed that the performance of models with short-range message passing strongly depends on the match between the effective interaction cutoff and the problem to be solved. The case of biodimers, demonstrates clearly that message passing cannot resolve interactions where no intermediate atoms are present. 
To study these cases, we perform additional experiments using \lorem, with and without long-range message passing, and with different cutoffs.


\begin{wrapfigure}[17]{l}{0.61\textwidth}
  \centering
    \begin{tikzpicture}[
        x = 1cm,
        y = 1cm,
        ]
        \node (figure) at (0,0) {\includegraphics[scale=0.6]{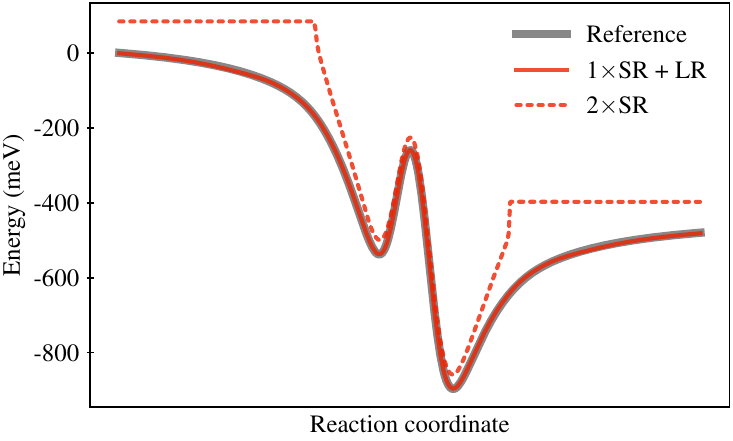} };
        \node () at (-1.6,1.0) {\includegraphics[scale=0.03]{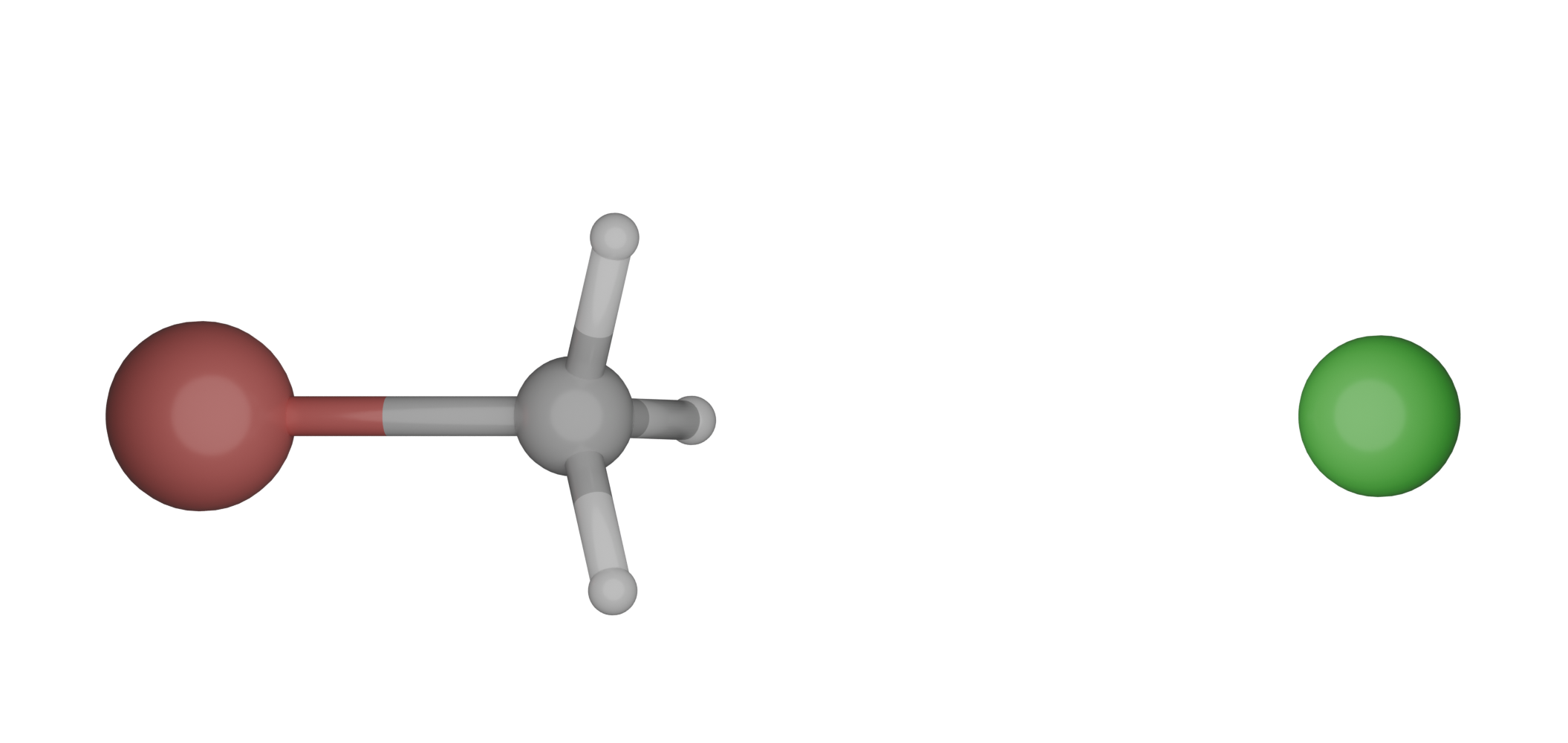} };
        \node () at (2.55,-0.9) {\includegraphics[scale=0.03]{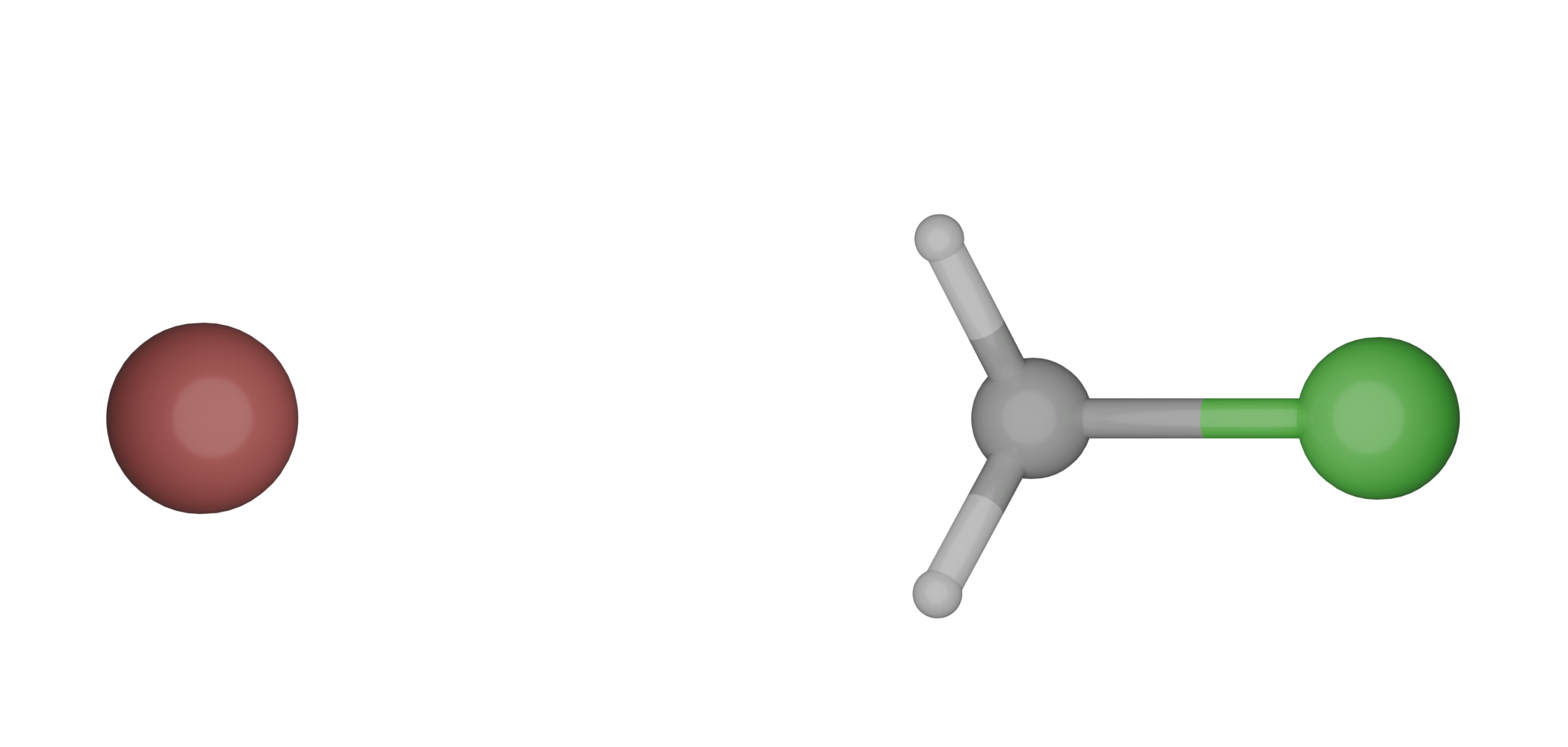} };
    \end{tikzpicture}
    \caption{Energy over the reaction coordinate for the nucleophilic substitution reaction $\ch{Cl^-} + \ch{H_3C-Br} \rightarrow \ch{Cl-CH_3} + \ch{Br^-}$; snapshots are shown as insets. }
    \label{fig:sn2}
\end{wrapfigure}

\paragraph{\sn reactions} Solving the benchmark task for the \sn reactions dataset requires long-range interactions, since it involves modeling intra-molecular interactions over distances exceeding the typical cutoffs used for \mlps. This is confirmed by \cref{fig:sn2}, which compares the performance of \lorem with and without long-range message passing. The former can reproduce the energy over the course of the reaction with excellent accuracy, including the tails as the reactants approach and separate. The latter, which does not include long-range interactions, cannot account for the tails and consequently predicts a constant once the molecules separate more than \qty{5}{\angstrom}.

\paragraph{Cumulene with different cutoffs} We probe the dependence of the performance of message-passing models on their hyperparameters by training a set of \lorem models with and without long-range message passing at different cutoffs (\qtylist{2.5;3;3.5}{\angstrom}) and numbers of short-range message passing steps (\numlist{1;2;3;4}). Small cutoffs are chosen to simulate the longer chain lengths of real biomolecular systems.
The results are shown in \cref{tab:cumulene-cutoffs} with additional plots in \cref{sub:apx-cumulene}: In all cases, models that include long-range message passing are able to resolve the angle. Short-range models, on the other hand, can only resolve the angle in certain combinations of hyperparameters: One message passing step is never sufficient. For two, a minimum cutoff of \qty{3.5}{\angstrom} is required. For three message passing steps, \qty{3.0}{\angstrom} is required.
Therefore, an effective cutoff substantially larger than the \qty{5.8}{\angstrom} distance between the central carbon atom in the chain and the hydrogen rotors is needed.
This is because \mpnns rely on the input graph structure (see \cref{fig:cumulene}A) for information flow: Below $\cutoff=\qty{2.6}{\angstrom}$, connections in the graph only extend to nearest neighbors, with the exception of the rotors and the second-to-last carbon atoms. Consequently, at least four message passing steps are required.
This example illustrates that short-range message passing is difficult to apply to this class of problems: Parameters have to be adapted to the chain length, with careful consideration of graph structure. In contrast, long-range message passing is robust, requiring no change in model hyper-parameters to solve this task.

\begin{table}
    \centering
    \caption{Ability of different \lorem models, with and without long-range message passing and with different numbers of short-range message passing steps, to solve the cumulene benchmark task. A tick (\yes) indicates yes, a cross (\no) indicates no. \Cref{fig:cumulene-cutoff} shows the full curves.}
    \begin{tabular}{r|cc cccc}
\toprule
              Cutoff  &  \makecell{$1\times$SR \\+ LR}  &  \makecell{$2\times$SR \\+ LR}  &           $1\times$SR  &           $2\times$SR  &           $3\times$SR  &           $4\times$SR \\ 
\midrule
\qty{2.5}{\angstrom}  &                  \yes  &                  \yes  &                   \no  &                   \no  &                   \no  &                  \yes \\ 
\qty{3.0}{\angstrom}  &                  \yes  &                  \yes  &                   \no  &                   \no  &                  \yes  &                  \yes \\ 
\qty{3.5}{\angstrom}  &                  \yes  &                  \yes  &                   \no  &                  \yes  &                  \yes  &                  \yes \\ 
\bottomrule
\end{tabular}

    \label{tab:cumulene-cutoffs}
\end{table}


\section{Discussion}

We introduced a simple yet effective equivariant long-range message passing scheme: Latent equivariant charges are predicted from local node features, and well-established techniques for evaluating inverse power-law potentials are used to efficiently compute long-range messages in a way that is convergent and well-defined for periodic systems. Building on this message-passing mechanism, we developed \lorem, a \mlp architecture that achieves consistently strong performance across benchmarks that require accurate \ML{long-range modeling}.

By construction, our model assumes that interactions decay asymptotically with distance. While it is therefore not suited for learning truly global representations that have no notion of locality, this limitation is largely theoretical for physical systems: Electrostatics dominates most long-range behavior in realistic systems, and even other effects typically do not extend over arbitrary distances.

A more practical limitation arises in non-periodic systems, where the cost of a naive long-range message evaluation scales as $\bigo{N^2}$. Although this is acceptable for small molecules and unit cells, this is a bottleneck for larger systems. However, linear-scaling methods for the evaluation of inverse power-law potentials are available: Fast multipole methods \citep{gr1987,pybbfmm} or multi-level summation \citep{hwpsss2015,bsucm2025pre}. Being able to leverage such methods is a key advantage of our proposed physics-inspired message-passing scheme.

We also investigated the capabilities of purely short-range message passing and found that, in many cases, it performs very well—even on datasets explicitly designed to require long-range interactions or charge equilibration. However, its success depends on matching the cutoff radius and number of message passing steps to the specific task, a process that can be both tedious and error-prone. More fundamentally, short-range methods cannot resolve interactions between distant atoms without intermediaries.
Our augmentation with long-range message passing overcomes these limitations, \CL{providing robust results across different tasks without requiring changes to model architecture or its hyperparameters (cutoff radius, number of message-passing steps, representation order). Training hyperparameters such as learning rate, optimizer, and loss weights are tuned per dataset, as is standard practice for \mlps}.

Finally, our results underscore a broader issue: the lack of challenging long-range benchmarks. Many current datasets are based on simplified, small-scale systems and fail to capture the complexity of real-world applications where long-range interactions are essential. \CL{Our deliberate focus on these benchmarks reflects the proof-of-concept nature of this work: On large, heterogeneous datasets, improvements in aggregate loss metrics can often be achieved by increasing the capacity of the short-range model, making it difficult to isolate and attribute gains to improved modeling of long-range physics \citep{hlhc2023}. The controlled benchmarks we consider allow a clearer mechanistic validation of the proposed equivariant Ewald message passing. Nevertheless, as a step towards larger-scale evaluation, we present results on the ADAPT silicon point-defect dataset \citep{dxzhrjk2025pre} in \cref{sec:apx-adapt}, where \lorem achieves competitive force accuracy and substantially better energy predictions compared to all baselines, using default model hyperparameters.}
Addressing \ML{the gap of challenging large long-range datasets} is a key direction for future work, which requires the development of application-oriented benchmarks.

\section*{Reproducibility statement}
\noindent
Code, data, configuration files, and trained models are available at \href{https://doi.org/10.5281/zenodo.17789350}{doi:10.5281/zenodo.17789350}. Scripts include data pre-processing, model training, model evaluation, and the creation of figures and tables in this work. Hyperparameters are also described in \cref{sec:apx-models}.

\bibliography{zotero, custom}
\bibliographystyle{tmlr}

\newpage
\appendix
\crefalias{section}{appendix}

\section{Equivariant modules in \lorem}
\label{sec:apx-modules}

As discussed in \cref{sec:background}, only certain operations can be applied to equivariant features $\tS$ without disrupting equivariance. We briefly discuss the operations used in \lorem; a more thorough discussion can be found in \citet{s2021,um2024}. All operations are implemented using the \texttt{e3x} library \citep{um2024}.
We work with spherical features $\tS$, dropping the atom index $i$, as operations are typically broadcast across $i$ or pairs of atoms.

$\tS$ is a three-dimensional tensor $\etS_{l,m,c}$ in which $l$ enumerates the order of the irreducible representation of $\sog$, $m$ the $2l+1$ components of that irreducible representation, and $c$ the channel (feature) dimension. For a fixed $l$ and $c$, $\tS_{l,:,c}$ can be thought of as a generalised Cartesian vector. Indeed, for $l=1$ it is a vector in three-dimensional space.

\paragraph{Prerequisites} A rotation $g \in \sog$ applied to all inputs causes a corresponding rotation of the spherical features $\tS$ expressed in terms of irreducible representations. For a set of features with order $l$, this rotation is represented by a matrix $\mR(g) \in \reals^{2l+1 \times 2l+1}$ and acts on these features through matrix multiplication along the $m$ index,
\begin{equation}
    \tS_{l,m,:} \longrightarrow_{g} \sum_{m'=-l}^{l} \emR_{m,m'}(g) \etS_{l,m',:} \, .
\end{equation}
In other word, the action of rotations on spherical features is a linear map.

\paragraph{Addition, Multiplication, Linear layers} Since the action of rotations is linear, adding two spherical features of equal $l$ together does not disrupt equivariance. By the same logic, any multiplication of spherical features that is broadcast along the $m$ index, i.e., that scales spherical features of order $l$ equally, is permissible.
Therefore, a linear layer can be applied to spherical features, provided it is only applied to the channel index, and broadcast along $m$. We therefore define a learned linear transformation as
\begin{equation}
    \text{Linear}(\tS_{l,:,c}) = \sum_{c'} \emW_{c,c'}^l \tS_{l,:,c'}
\end{equation}
with per-$l$ learned weights $\mW^l$. A bias term could be applied to $l=0$, i.e., the scalar part of the spherical features, but we choose not to.

\paragraph{Tensor products} Two spherical features of order $l_1$ and $l_2$ can be combined into a new spherical feature with $l_3$ using a specialized tensor product
\begin{equation}
    \text{Tensor}(\tS_{l_1,m_1,:}, \tQ_{l_2,m_2,:}) = \tU_{l_3,m_3,:}
    = \evw^{l_1,l_2,l_3}_{:} \sum_{m_1,m_2} \etC_{m_1,m_2,m_3}^{l_1,l_2,l_3} \tS_{l_1,m_1,:} \tQ_{l_2,m_2,:}
\end{equation}
$\tC^{l_1,l_2,l_3}$ are the Clebsch-Gordan coefficients, and $\vw^{l_1,l_2,l_3} \in \reals^c$ is a per-channel weight vector for a given combination of $l_1,l_2,l_3$. This tensor product can be carried out for every valid combination $|l_1 - l_2| \leq l_3 \leq |l_1 + l_2|$.
In \lorem, unless specified otherwise, tensor products are only carried out to a fixed maximum dimension $\lmax$, which all spherical features have in common, not the maximum possible one. Some operations, for example the preparation of the equivariant message-passing block, perform a tensor product only to a specified target order.
If both inputs of the tensor product are the same feature, we call the operation a self-tensor product. This operation increases the body-order of the representation \citep{dbcdevo2022,bksoc2022}.

\paragraph{Spherical Norm} To predict invariant quantities, we require a way to extract invariant information from spherical, equivariant, features. One way to do achieve this is a tensor product to target order $l=0$. Another, which is used in \lorem, is to take the norm of each irreducible representation
\begin{equation}
    \text{Norm}(\tS) = \sqrt{(2l+1)^{1/2} \sum_m \etS_{l,m,c}^2 } \, .
\end{equation}
We found empirically that the prefactor $(2l+1)^{1/2}$ helps reduce variance across $l$.
As opposed to a tensor product to $l=0$, this operation, inspired by the power spectrum \citep{bkc2013}, keeps the norms per $l$ separate; a tensor product would linearly combine all $l$ into one.

\section{Invariance of \lorem}
\label{sec:ap_equiv}

As explained in \cref{sec:lrmp,sec:apx-modules}, the potential at each atom, $\tV_i$ is equivariant because the long-range message passing step consists of equivariant operations: Addition and multiplication with a factor shared per $l$. In practical implementations of Ewald summation in periodic systems, there is one extra step: To prevent divergences, the total charge must be zero, which can be done either by simply subtracting the sum of the total charge from each charge, or equivalently, by subtracting an analytical correction from the potentials. Both approaches are equivalent, and since summation is equivariant, also do not break invariance. For this reason, the total procedure retains equivariance.

To numerically confirm these considerations, we compute the mean absolute error of energy predictions over a degree $L=3$ Lebedev grid of rotations, plus inversions, with respect to the unrotated case, for the first \num{5} structures of the MgO surface validation set. The errors are in line with precision expectations: For single precision, \qty{1.031e-6}{meV} and for double precision \qty{5.913e-15}{meV}.

\section{Dataset Construction}
\label{sec:ap_datasets}

\paragraph{\ch{MgO} surface}
The \ch{MgO} surface dataset from \cite{kfgb2021} is used without modification. A representative snapshot of the unit cell is shown in \cref{fig:aumgo+nacl}A. The dataset is built from four configuration types, each derived from a distinct initial structure:
\begin{enumerate}
\item A pure \ch{MgO} surface with the \ch{Au2} dimer oriented perpendicular to the surface,
\item A pure \ch{MgO} surface with the \ch{Au2} dimer oriented parallel to the surface,
\item An Al-doped \ch{MgO} surface with the \ch{Au2} dimer perpendicular to the surface, and
\item An Al-doped \ch{MgO} surface with the \ch{Au2} dimer parallel to the surface.
\end{enumerate}
Each of these initial configurations was first geometry optimized. For the two perpendicular, \scare{non-wetting}, cases (1 and 3), the distance between the lower \ch{Au} atom and the \ch{O} atom directly beneath it was systematically varied, displacing the \ch{Au2} dimer as a whole. From these distance-dependent samples, a subset was randomly selected.
To introduce structural diversity, Gaussian noise was applied to each configuration: a standard deviation of \qty{0.02}{\angstrom} for atoms in the \ch{MgO} substrate and \qty{0.1}{\angstrom} for the gold cluster. After perturbation, \num{1250} structures were selected from each of the four configuration types, yielding a total of \num{5000} samples. For these, energies and forces were computed using the FHI-aims code \citep{aaazzz2025pre} and the Perdew–Burke–Ernzerhof (PBE) functional \citep{pbe1996}.
A random \num{90}/\num{10} train–validation split was used, as reported in \cite{kfgb2021}. Additional energy–distance curves with equal spacing were constructed for the perpendicular cases and are shown in \cref{fig:aumgo+nacl}B.

\paragraph{\ch{NaCl} cluster}
The \ch{NaCl} cluster dataset from \cite{kfgb2021} is used without modification.
First, the \ch{Na9Cl8^+} cluster was optimized in vacuum. A snapshot is shown in the top panel of \cref{fig:aumgo+nacl}C. From this geometry, the \ch{Na} atom farthest from all other \ch{Na} atoms was removed, yielding the \ch{Na8Cl8^+} cluster shown in the bottom panel of \cref{fig:aumgo+nacl}C.
From these two structures, additional configurations were created by varying the distance between a selected pair of \ch{Na} atoms along the line connecting them. In \cref{fig:aumgo+nacl}C, this moving atom is illustrated by transparent copies along its trajectory.
Training datasets for both \ch{Na9Cl8^+} and \ch{Na8Cl8^+} were constructed by randomly sampling configurations along these trajectories. Gaussian noise with a standard deviation of \qty{0.05}{\angstrom} was applied to the atomic coordinates, resulting in \num{2500} perturbed structures for each molecule and a total of \num{5000} instances. Energies and forces were computed using the FHI-aims code and the PBE functional.
A random \num{90}/\num{10} train–validation split was used, as reported by \cite{kfgb2021}. Additional energy–distance curves with equal spacing were constructed and are shown in \cref{fig:aumgo+nacl}D.

\paragraph{Cumulenes}
The cumulene dataset from \cite{ucsgpstm2021} is used without modification. The geometry of the linear molecule is shown in \cref{fig:cumulene}A. The dataset contains \num{4500} randomly sampled cumulene structures with nine carbon atoms, divided into training, validation, and test sets with \num{2000}/\num{500}/\num{2000} instances, respectively.
The energy curves shown in \cref{fig:cumulene}B are based on a controlled subset in which only one terminal carbon atom is rotated, while the rest of the molecule remains fixed. A visualization of this rotational motion is provided in the inset of \cref{fig:cumulene}B.

\paragraph{Biodimers}
The Biodimers dataset from~\cite{hlhc2023, burns2017biofragment} is used without modification. It consists of \num{2291} relaxed organic sidechain–sidechain fragments, including small molecules such as ethanol, acetamide, and others.
Based on molecular properties, the dataset is divided into six categories: each molecule is classified as either polar, apolar, or charged, resulting in the following dimer types—polar–polar (PP), polar–apolar (PA), charged-polar (CP), apolar–apolar (AA), charged-apolar (CA), and charged–charged (CC). A representative CC dimer is shown in \cref{fig:bio_dimers}A.
The initial separation between molecules reflects their positions in protein structures. From this configuration, the intermolecular distance is incrementally increased up to \qty{15}{\angstrom}, resulting in a total of \num{29783} dimer instances.
Instances where the final separation exceeds the initial distance by more than \qty{4}{\angstrom} (\num{13743} samples) are designated as the test set. The remaining \num{16040} instances form the training set. Energies and forces were computed using the FHI-aims code and the HSE06 hybrid functional. For each of the six dimer types, a random \num{80}/\num{20} train–validation split was applied. The resulting subsets were then merged into a single training set and a single validation set.

\paragraph{\sn reactions}
The \sn reactions dataset from \cite{um2019, fcmu2024} is used without modification. It contains molecular structures in vacuum that model nucleophilic substitution (\sn) reactions. The dataset includes molecules of the types \ch{XCH3Y^-}, \ch{CH3X}, \ch{HX}, \ch{CHX}, \ch{CH2X}, \ch{XY}, \ch{X}, and \ch{Y}, for all possible combinations of \( \ch{X}, \ch{Y} \in \{\ch{F}, \ch{Cl}, \ch{Br}, \ch{I}\} \). A representative reaction coordinate with corresponding snapshots is shown in \cref{fig:sn2}. Additional species such as \ch{H2}, \ch{CH2}, and \ch{CH3} are also included in the dataset.
Training configurations were generated using ab initio molecular dynamics simulations at \qty{5000}{\kelvin}, with a time step of \qty{0.1}{\femto\second}. Full computational details, including the level of theory, are provided in the original publications. The dataset is randomly split into \num{405000} training, \num{5000} validation, and \num{42708} test samples.

\section{Model description and training setup}
\label{sec:apx-models}

For all models, training parameters, and in some cases, model parameters, vary slightly between datasets. Where we were unable to use models from previous work, we tuned parameters to minimize the error on the validation set,\footnote{We note that minimizing the validation error does not always yield the best qualitative agreement: In the cumulene dataset, it does not readily correlate with ability to predict the dihedral curve in \cref{fig:cumulene} -- while all models below a certain error threshold are able to resolve the task, the qualitative agreement varies.} aiming to perform a similar number of experiments, on the order of ten, per model and experiment. With this, we aim to report results that reflect a practical degree of hyper-parameter tuning.

\paragraph{\lorem}
For all experiments, except the ones where variations were tested, the same \lorem model  architecture (\cref{sec:model}) was used: A cutoff radius of \qty{5}{\angstrom}, $\lmax=6$ for spherical features and $\lrlmax=2$ for the long-range message passing, \num{128} scalar features, \num{8} channels for spherical features, and \num{32} radial basis functions. We perform only the initial short-range message passing step; performing additional steps typically increases accuracy but hinders comparison of long-range expressivity. This model has \num{1021198} learnable parameters. It was implemented using the \texttt{e3x} library \citep{um2024}.
Training is performed entirely in \texttt{float32} precision; while we find that reduced precision has only a minor effect on training dynamics, it can significantly alter validation and test results.

Training parameters vary slightly between datasets; the exact parameters can be found in \cref{tab:aeres-params}. We report errors for the better out of two runs with different seeds; the presented conclusions hold for both models.

In all cases, except NaCl\footnote{Here, exponential learning rate decay was employed.}, the learning rate was decayed linearly after 10 epochs, from the indicated starting value to \num{1e-6}; the ADAM \citep{kb2017} and LAMB \citep{ylrhkbsdkh2019} optimizers were used. The loss function was a simple mean squared error, with the energy residuals normalized by number of atoms. The squared residuals were averaged over the whole batch, including over atoms and components in the case of forces, and then summed and weighted with a factor. The checkpoint with the lowest summed $R^2$ of energy and forces, evaluated on the validation set, was used for experiments. Training times are given for the entire run, not the time until the best checkpoint.

For NaCl and AuMgO, prior work \citep{kkzc2024,kfgb2021} reports error directly on the validation set, as the benchmark was originally intended as an overfitting exercise. We follow this practice, but verify in \cref{sec:apx-valset} that, due to the statistical uniformity of the dataset, there is no significant difference to tuning hyper-parameters systematically on an inner train/validation split, keeping the data used for \cref{tab:metrics} held out.

\begin{table}[]
    \centering
    \begin{tabular}{r|rrrrrrr}
    \toprule
        Dataset & Optimizer & Initial LR & Epochs & Batch size & $E$ weight & $\F$ weight & Time \\
    \midrule
        MgO surface & ADAM &\num{4e-4} & \num{4000} & \num{32} & \num{1000} & \num{1.0} & \qty{17}{h} \\
        Biodimers & ADAM & \num{1e-4} & \num{4000} & \num{32} & \num{0.5} & \num{0.5} & \qty{42}{h} \\
        Cumulene & LAMB & \num{1e-3} & \num{2000} & \num{32} & \num{0.5} & \num{0.5} & \qty{2}{h} \\
        NaCl cluster & ADAM &  \num{1e-3} & \num{8000} & \num{64} & \num{0.5} & \num{0.5} & \qty{7}{h} \\
        \sn reactions & ADAM &  \num{1e-4} & \num{500} & \num{32} & \num{0.5} & \num{0.5} & \qty{17}{h} \\
    \bottomrule
    \end{tabular}
    \caption{Training settings for \lorem for different datasets. Training times are given for a single Nvidia H100 SXM5 GPU.}
    \label{tab:aeres-params}
\end{table}

\paragraph{\mace}
The \mace \mlp (\cite{bksoc2022}) is an equivariant message-passing neural network as described in \cref{sec:background}. It is used with the standard setting of two message-passing steps, i.e., an effective cutoff of \qty{10}{\angstrom}.
Training generally used default hyperparameters, with some adjustments made on hyperparameters related to training dynamics. Specifically, the cumulene system was trained using the default energy-to-forces loss weight ratio of 1:100 and '128x0e + 128x1o + 128x2e' hidden irreps. All other systems were trained using a 1:10 ratio and '128x0e + 128x1o' hidden irreps.

All experiments, except those involving cumulenes and the NaCl cluster, used the SWA protocol, which swaps the loss weights between energy and forces at a specified epoch. This epoch was set where the energy RMSE plateaued, with the subsequent training continued until loss saturation. All models were trained with $l_{\text{max}} = 2$, a batch size of 32, and utilized cuEquivariance acceleration.

\paragraph{\pet}
\pet (\cite{pc2023}) is an unconstrained transformer model, consisting of multiple edge-to-edge transformer layers within local neighborhoods followed by message passing. Similar to \mace, it is also used with two message-passing steps and an effective cutoff of \qty{10}{\angstrom}. To ensure approximate rotational invariance, it is trained with data augmentation. No inference-time symmetrization is used in our experiments.

Unless otherwise specified, we used default hyperparameters (cutoff radius of \qty{5}{\angstrom} with cosine smoothing over the outermost \qty{0.2}{\angstrom}, $d_{\text{PET}}=128$, $d_{\text{head}}=128$, $d_{\text{feedforward}}=512$, with 8 heads per attention layer, and 2 attention layers per GNN layer), with some adjustments made related to training dynamics. Models were trained using an epoch-based scheduler, which halved the learning rate after 250 epochs. This applies to all datasets except biodimers. For biodimers, a \texttt{ReduceLROnPlateau} scheduler was used instead, reducing the learning rate by 20\% if the loss did not improve for 100 consecutive epochs. To improve training stability, biodimers training also employed gradient clipping with a maximum gradient norm of 5. Every training run included 10 warmup epochs, during which the learning rate was linearly increased from zero to the preset learning rate of $1 \times 10^{-4}$.

For the biodimers and NaCl cluster datasets, the targets were normalized by their standard deviation in the training set. The energy-to-forces loss weight ratio was set to 1:1 for biodimers and MgO surface datasets, while the NaCl cluster dataset used a ratio of 1:10.

For cumulene, PET was trained for \num{20100} epochs with a batch size of \num{16}, equal energy and forces weight, and a maximum learning rate of \num{2e-4}. The learning rate was increased linearly from \num{0} over 100 epochs, and then reduced by 10\% every 500 epochs over the entire training run. Architecturally, \num{4} attention layers and cosine cutoff function was employed to resolve the cumulenes. We found that increasing the number of attention layers, and adapting the cutoff function, were critical to resolve this benchmark task.

\paragraph{\cace}
The \cace model was originally presented in~\cite{c2024}, with an additional modification introducing a long-range component in~\cite{c2025}. Similar to \mace, \cace is an equivariant message-passing neural network. After short-range message passing, invariant scalar pseudo-charges are predicted and passed to the long-range part of Ewald summation; the resulting energy contribution is added to the energy predictions of the short-range message-passing model. We use this model with the recommended setting of one message-passing step and, depending on the system, the following corresponding effective cutoff radii: \qty{5.5}{\angstrom} for the \ch{MgO} surface, \qty{5.29}{\angstrom} for the \ch{NaCl} cluster—these are the settings used in the original paper \cite{kkzc2024}—and we chose a cutoff of \qty{5}{\angstrom} for biodimers and cumulene.

For the \ch{MgO} surface and \ch{NaCl} cluster datasets, models were taken from~\cite{kkzc2024}. For the cumulenes and biodimers datasets, the hyperparameters were as follows: 6 Bessel radial functions, $c = 8$, $l_{\text{max}} = 3$, $\nu_{\text{max}}$, $N_{\text{embedding}} = 2$, one message-passing layer, one-dimensional hidden variable, $\sigma = 1$, and $dl = 2$. Training followed the example in the original paper: the first 200 epochs used an energy loss weight of 0.1 and a forces loss weight of 1000, after which the energy loss weight was changed to 1, 10, and 1000 every 100 epochs, yielding a total of 500 epochs. The learning rate was $5 \times 10^{-3}$, with a step learning rate schedule decreasing it by a factor of 2 every 20 steps.

\paragraph{\fgnn}
The \fgnn model was designed to tackle the benchmarks introduced by \cite{kfgb2021}, and introduced in that work. It consists of a shallow neural network acting on rotationally invariant features, predicting both a local energy contribution and an electronegativity, which is then used in a charge equilibration procedure that globally redistributes charges to minimize an energy expression. This process can be seen as a physics-inspired long-range message passing scheme iterated until a fixed point is reached. 
As the model requires charge labels to train, which are not available for all datasets, we did not train this model for our experiments but instead include results from \citet{kfgb2021}, which are only available for the \ch{NaCl} cluster and \ch{MgO} surface datasets. The cutoffs used in the original paper are as follows: \qty{4.23}{\angstrom} for the \ch{MgO} surface and \qty{5.29}{\angstrom} for the \ch{NaCl} cluster.

\CL{
\paragraph{SpookyNet}
SpookyNet \citep{ucgssm2021} is an equivariant \mlp that predicts scalar partial charges and nuclear dipoles, which are used to compute long-range electrostatic and dispersion corrections via Ewald summation. Additionally, the model includes global attention without geometric information. The model has approximately \qty{3}{M} parameters \citep{bmc2023}. Results for the \ch{MgO} surface and \ch{NaCl} cluster datasets are taken from \citet{kfgb2021}.
}

\clearpage
\section{Runtime Benchmark}
\label{sec:apx-benchmark}

\CL{To estimate the runtime overhead of the long-range block and compare the Ewald and particle-mesh Ewald (PME) implementations, we benchmarked \lorem on supercells of three physical crystal structures: \ch{NaCl}, \ch{CsCl}, and \ch{Cu_2O}. For each crystal, we constructed supercells of increasing size and ran predictions (energy and forces) for \lorem models with and without the long-range block, using both Ewald and PME for the long-range evaluation. All benchmarks were repeated ten times and averaged, and executed on a single NVIDIA H100 GPU.}

\begin{figure}
    \centering
    \CL{\includegraphics[scale=0.65]{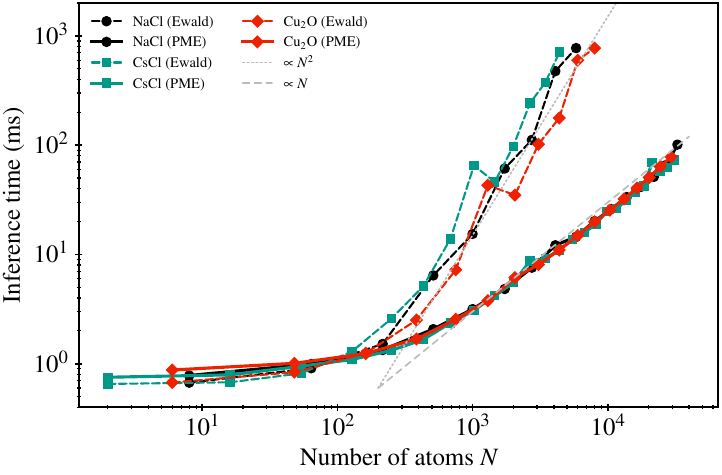}}
    \caption{\CL{Runtime scaling of \lorem with Ewald (dashed) and PME (solid) long-range implementations for \ch{NaCl}, \ch{CsCl}, and \ch{Cu_2O} supercells. Reference scaling lines ($\propto N^2$ and $\propto N$) are shown for comparison.}}
    \label{fig:benchmark-scaling}
\end{figure}

\CL{The scaling behavior is shown in \cref{fig:benchmark-scaling}. Ewald summation scales quadratically with system size and runs out of memory beyond approximately \num{8000} atoms; PME scales roughly linearly and extends to over \num{30000} atoms. At \num{4096} atoms, PME is approximately $40\times$ faster than Ewald. Detailed timings, including the fraction of total runtime spent in the long-range block, are reported in \cref{tab:benchmarks-nacl,tab:benchmarks-cscl,tab:benchmarks-cu2o}. With Ewald, the LR fraction grows from around \qty{30}{\percent} at small sizes to over \qty{95}{\percent} at large sizes. With PME, the LR fraction remains roughly constant and depends on the density of the crystal: \ch{Cu_2O} (the densest, $\sim 35$ neighbors per atom) shows around \qty{25}{\percent}, while \ch{CsCl} (the least dense, $\sim 14$ neighbors) shows around \qty{60}{\percent}. Overall, performance with PME on the order of single-digit \unit{\micro\second}/atom is in line with expectations for modern \mlps.}

\begin{table}
    \caption{\CL{Runtime of \lorem with Ewald and PME long-range implementations for \ch{NaCl} supercells of varying size $N$. LR denotes the time spent in the long-range block alone.}}
    \centering
    \CL{\begin{tabular}{r | r r r | r r r}
\toprule
  \multicolumn{1}{c}{} & \multicolumn{3}{c}{Ewald} & \multicolumn{3}{c}{PME} \\ 
           $N$  &      SR+LR (ms)  &         LR (ms)  &         LR (\%)  &      SR+LR (ms)  &         LR (ms)  &         LR (\%) \\ 
\midrule
             8  &       \num{0.7}  &       \num{0.2}  &      \num{32.4}  &       \num{0.8}  &       \num{0.3}  &      \num{41.9} \\ 
            64  &       \num{0.9}  &       \num{0.4}  &      \num{45.4}  &       \num{1.0}  &       \num{0.5}  &      \num{50.0} \\ 
           512  &       \num{6.4}  &       \num{5.3}  &      \num{82.6}  &       \num{2.1}  &       \num{1.0}  &      \num{46.0} \\ 
          1728  &      \num{61.2}  &      \num{58.4}  &      \num{95.5}  &       \num{4.8}  &       \num{2.1}  &      \num{42.6} \\ 
          4096  &     \num{475.8}  &     \num{469.8}  &      \num{98.7}  &      \num{12.1}  &       \num{6.1}  &      \num{50.5} \\ 
          8000  &              --  &              --  &              --  &      \num{20.2}  &       \num{8.9}  &      \num{43.8} \\ 
         17576  &              --  &              --  &              --  &      \num{41.5}  &      \num{16.9}  &      \num{40.7} \\ 
         32768  &              --  &              --  &              --  &     \num{101.2}  &      \num{51.6}  &      \num{51.0} \\ 
\bottomrule
\end{tabular}
}
    \label{tab:benchmarks-nacl}
\end{table}

\begin{table}
    \caption{\CL{Runtime of \lorem with Ewald and PME long-range implementations for \ch{CsCl} supercells of varying size $N$.}}
    \centering
    \CL{\begin{tabular}{r | r r r | r r r}
\toprule
  \multicolumn{1}{c}{} & \multicolumn{3}{c}{Ewald} & \multicolumn{3}{c}{PME} \\ 
           $N$  &      SR+LR (ms)  &         LR (ms)  &         LR (\%)  &      SR+LR (ms)  &         LR (ms)  &         LR (\%) \\ 
\midrule
             2  &       \num{0.7}  &       \num{0.3}  &      \num{42.7}  &       \num{0.8}  &       \num{0.4}  &      \num{50.2} \\ 
            54  &       \num{0.8}  &       \num{0.4}  &      \num{48.1}  &       \num{0.9}  &       \num{0.5}  &      \num{55.4} \\ 
           432  &       \num{5.1}  &       \num{4.3}  &      \num{84.5}  &       \num{1.7}  &       \num{0.9}  &      \num{52.4} \\ 
          1458  &      \num{46.0}  &      \num{44.3}  &      \num{96.2}  &       \num{4.2}  &       \num{2.5}  &      \num{58.8} \\ 
          4394  &     \num{710.1}  &     \num{705.7}  &      \num{99.4}  &      \num{11.0}  &       \num{6.7}  &      \num{60.6} \\ 
          8192  &              --  &              --  &              --  &      \num{19.0}  &      \num{11.2}  &      \num{59.2} \\ 
         18522  &              --  &              --  &              --  &      \num{42.5}  &      \num{25.5}  &      \num{60.0} \\ 
         31250  &              --  &              --  &              --  &      \num{72.9}  &      \num{44.2}  &      \num{60.7} \\ 
\bottomrule
\end{tabular}
}
    \label{tab:benchmarks-cscl}
\end{table}

\begin{table}
    \caption{\CL{Runtime of \lorem with Ewald and PME long-range implementations for \ch{Cu_2O} supercells of varying size $N$.}}
    \centering
    \CL{\begin{tabular}{r | r r r | r r r}
\toprule
  \multicolumn{1}{c}{} & \multicolumn{3}{c}{Ewald} & \multicolumn{3}{c}{PME} \\ 
           $N$  &      SR+LR (ms)  &         LR (ms)  &         LR (\%)  &      SR+LR (ms)  &         LR (ms)  &         LR (\%) \\ 
\midrule
             6  &       \num{0.7}  &       \num{0.3}  &      \num{42.2}  &       \num{0.9}  &       \num{0.5}  &      \num{55.6} \\ 
            48  &       \num{0.8}  &       \num{0.4}  &      \num{42.8}  &       \num{1.0}  &       \num{0.5}  &      \num{52.5} \\ 
           384  &       \num{2.5}  &       \num{1.5}  &      \num{58.6}  &       \num{1.7}  &       \num{0.6}  &      \num{38.4} \\ 
          1296  &      \num{43.0}  &      \num{40.3}  &      \num{93.9}  &       \num{3.8}  &       \num{1.1}  &      \num{30.4} \\ 
          4374  &     \num{177.1}  &     \num{169.1}  &      \num{95.4}  &      \num{11.0}  &       \num{2.9}  &      \num{26.5} \\ 
          7986  &     \num{773.1}  &     \num{759.0}  &      \num{98.2}  &      \num{19.9}  &       \num{5.7}  &      \num{28.7} \\ 
         16464  &              --  &              --  &              --  &      \num{40.5}  &      \num{11.3}  &      \num{27.8} \\ 
         29478  &              --  &              --  &              --  &      \num{78.1}  &      \num{19.6}  &      \num{25.1} \\ 
\bottomrule
\end{tabular}
}
    \label{tab:benchmarks-cu2o}
\end{table}

\clearpage
\section{Ablations of LR block and $l$}
\label{sec:apx-ablations}

We repeated the experiments on the \ch{MgO} surface, \ch{NaCl} cluster, cumulene, and biodimers with \lorem models with $l=0$ and $l=1$, as well as without the long-range message-passing block. The parameter counts for no LR and $l=0,1,2$ (in that order) are  $839129$, $1020300$, $1020749$, and $1021198$.

\begin{figure}
  \centering
    \begin{tikzpicture}[
        x = 1cm,
        y = 1cm,
        ]
        \node (figure) at (4,0) {\includegraphics[scale=0.65]{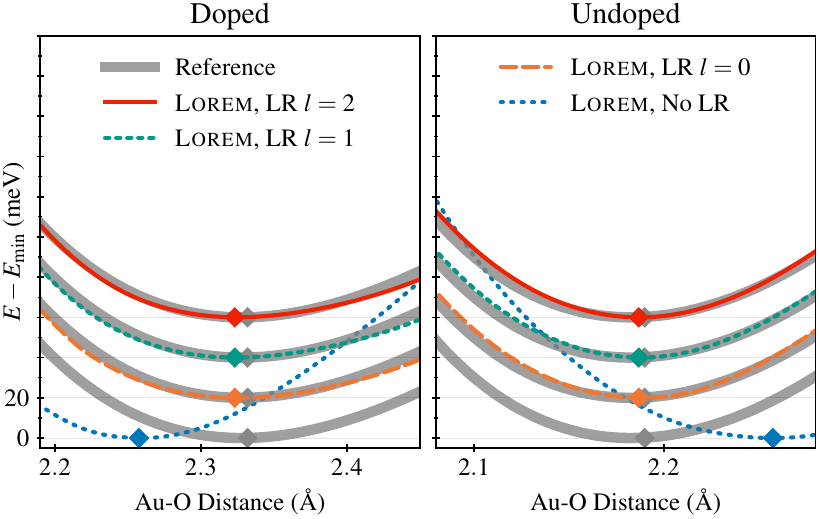}};
        \node (render) at (-3.2,0.1) {\includegraphics[scale=0.18]{figures/aumgo_render_small.pdf}};
        \node (l) at (-4.80, 0.20) {$\sim\hspace{-0.5mm}\qty{12}{\angstrom}$};
        \node (d) at (-2.35, 1.4) {$d$};
        \node (w) at (-4.70, 2.35) {non-wetting};
        \node (nw) at (-4.40, 2.00) {wetting};
        \node (dp) at (-4.80, -1.6) {dopant};
        \node[label] () at (-5, 2.9) {$\mathbf{A}$};
        \node[label] () at (-0.5, 2.9) {$\mathbf{B}$};
        \node (figure) at (4,-5.25) {\includegraphics[scale=0.65]{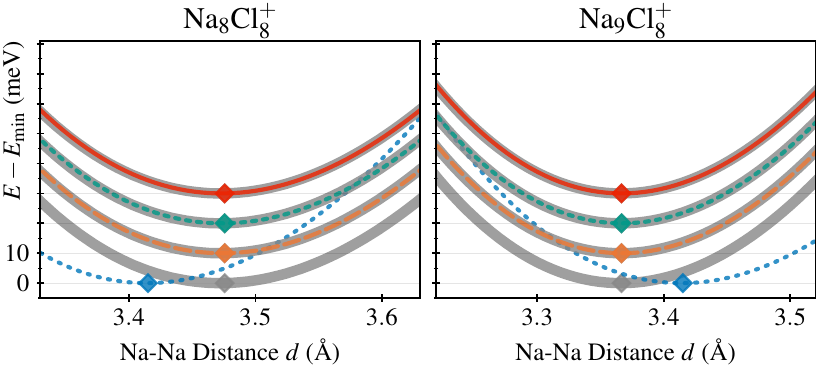}};
        \node (render) at (-3.2,0.1-5.25) {\includegraphics[scale=0.18]{figures/nacl_render_small.pdf}};
        \node (l) at (-3.15, 1.50-5.25) {$\sim\hspace{-0.5mm}\qty{21}{\angstrom}$};
        \node (d) at (-4.65, -1.50-5.25) {$d$};
        \node[label] () at (-5, 2.2-5.45) {$\mathbf{C}$};
        \node[label] () at (-0.5, 2.2-5.45) {$\mathbf{D}$};
    \end{tikzpicture}
    \caption{
    ($\mathbf{A}$) \ch{Au_2} dimer on \ch{MgO} surface, showing both wetting and non-wetting geometries, as well as the \ch{Al} dopant. ($\mathbf{B}$) Energy over distance $d$ for the non-wetting geometry for the doped and undoped surface. The minima are indicated with a diamond symbol; the reference energy curve is drawn in grey. Offsets are added to distinguish the curves and the value at the minimum is subtracted.
    ($\mathbf{C}$) \ch{Na_9Cl_8^+} (top) and \ch{Na_8Cl_8^+} (bottom) cluster, the moving atom is marked with transparent copies of itself, and the distance of interest is labeled with $d$.
    ($\mathbf{D}$) Energy over distance for both clusters.
    }
    \label{fig:ablation-aumgo+nacl}
\end{figure}

\begin{figure}
  \centering
    \includegraphics[scale=0.65]{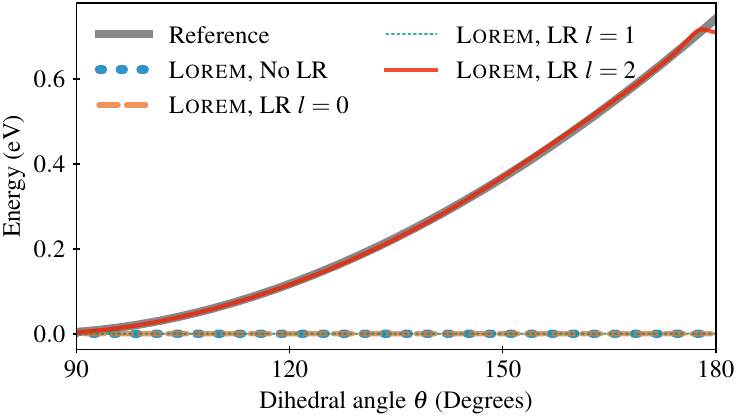}
    \caption{
    Energy profile over a \qty{90}{\degree} rotation of one rotor. The minimum value of each curve is subtracted before plotting.
    }
    \label{fig:ablation-cumulene}
\end{figure}

\begin{figure}
  \centering
    \begin{tikzpicture}[
        x = 1cm,
        y = 1cm,
        ]
        \node (figure) at (0.0,0) {\includegraphics[scale=0.63]{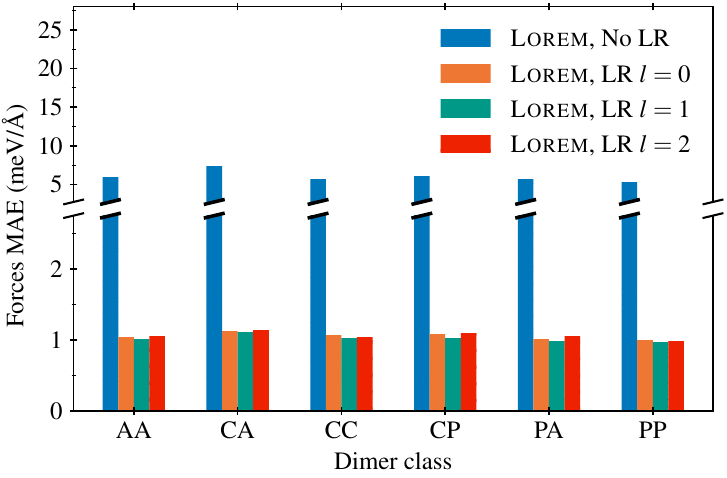} };
    \end{tikzpicture}
    \caption{
    Mean absolute error on forces for different models on the different dimer classes: Apolar-apolar (AA), charge-apolar (CA), charge-charge (CC), charge-polar (CP), polar-apolar (PA), polar-polar (PP).
    }
    \label{fig:ablation-bio_dimers}
\end{figure}

Results can be seen in \cref{tab:ablation-metrics,fig:ablation-aumgo+nacl,fig:ablation-cumulene,fig:ablation-bio_dimers}.
Generally, higher $l$ improve error metrics.
For experiments where only scalar long-range interactions are required (\ch{MgO} surface, \ch{NaCl} cluster), higher $l$ do not improve qualitative agreement.
The cumulene example, which requires access to relative orientation between rotors, requires equivariant long-range interactions with $l=2$ to be resolved.
Removing the long-range message-passing block significantly increases error and renders the model unable to solve all the benchmark tasks.

\begin{table}
    \caption{Root mean squared errors for energy $E$ and forces $\F$ for datasets used in Sec. 6.1 of the main text, for different \lorem variants. See Tab. 1 of the main text for details.}
    \centering
    \begin{tabular}{r | r r r r}
\toprule
                 Dataset  &  \makecell{\textsc{Lorem} \\No LR}  &  \makecell{\textsc{Lorem} \\LR $l=0$}  &  \makecell{\textsc{Lorem} \\LR $l=1$}  &  \makecell{\textsc{Lorem} \\LR $l=2$} \\ 
\midrule
\ch{MgO} surface\hspace{3mm} $E$ (meV/at)  &               \num{2.234}  &               \num{0.063}  &      \textbf{\num{0.062}}  &               \num{0.064} \\ 
{\small (Validation set)}\hspace{3mm} $\F$ (meV/Å)  &              \num{61.487}  &               \num{5.870}  &               \num{5.284}  &      \textbf{\num{4.076}} \\ 
\midrule
\ch{NaCl} cluster\hspace{3mm} $E$ (meV/at)  &               \num{1.582}  &               \num{0.113}  &      \textbf{\num{0.111}}  &               \num{0.112} \\ 
{\small (Validation set)}\hspace{3mm} $\F$ (meV/Å)  &              \num{50.110}  &               \num{1.243}  &      \textbf{\num{1.062}}  &               \num{1.155} \\ 
\midrule
Biodimers\hspace{3mm} $E$ (meV/at)  &               \num{8.259}  &               \num{0.302}  &               \num{0.329}  &      \textbf{\num{0.222}} \\ 
\hspace{3mm} $\F$ (meV/Å)  &              \num{16.452}  &               \num{1.677}  &               \num{1.725}  &      \textbf{\num{1.646}} \\ 
\midrule
Cumulene\hspace{3mm} $E$ (meV/at)  &              \num{16.203}  &              \num{15.793}  &               \num{5.576}  &      \textbf{\num{3.309}} \\ 
\hspace{3mm} $\F$ (meV/Å)  &             \num{126.000}  &             \num{133.092}  &              \num{85.956}  &     \textbf{\num{50.084}} \\ 
\bottomrule
\end{tabular}

    \label{tab:ablation-metrics}
\end{table}

\clearpage
\section{Additional results for cumulene}

\label{sub:apx-cumulene}

\begin{figure}
    \centering
    \includegraphics[scale=0.6]{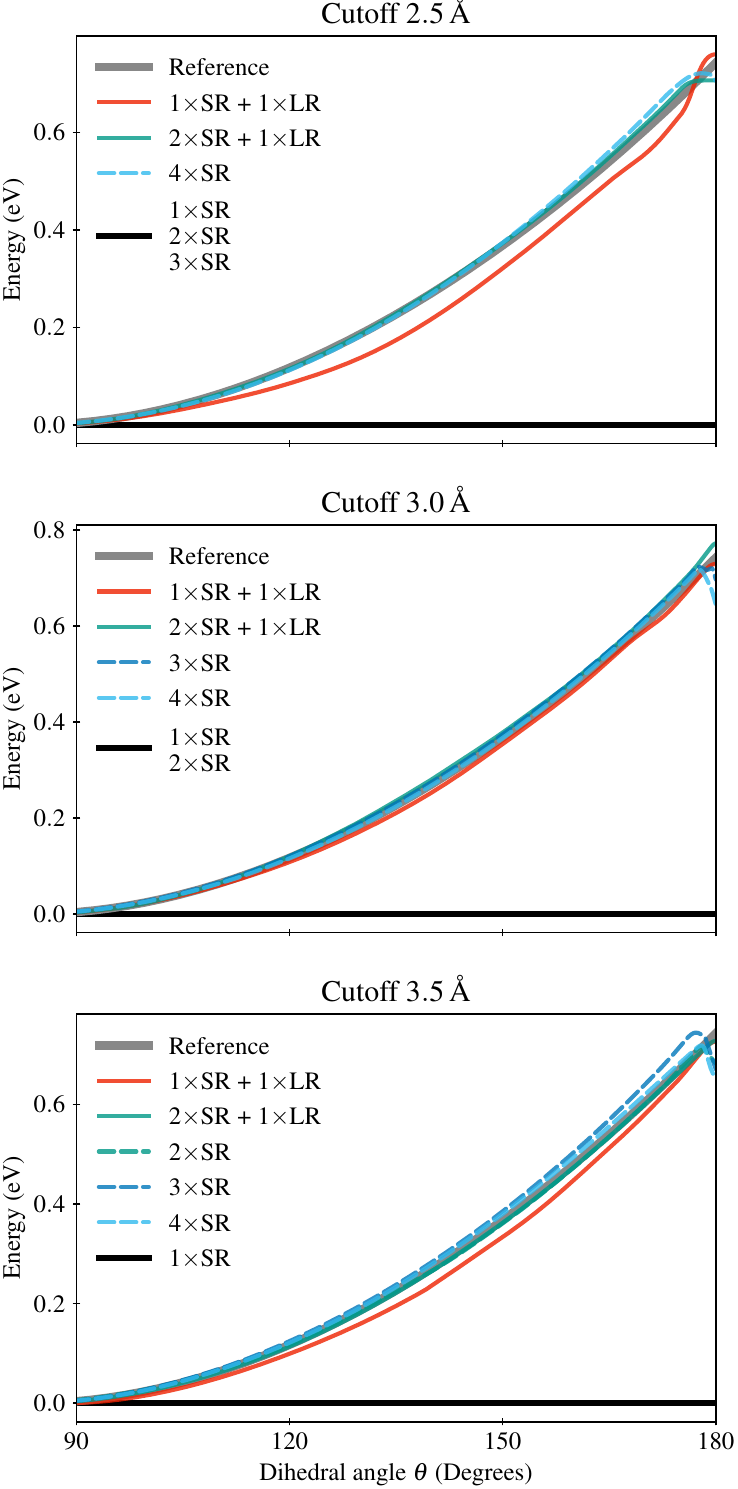}
    \caption{Rotational profile of cumulene for variations of the \lorem model with different $\cutoff$, different numbers of short-range message passing steps, and with and without long-range message passing. All curves that are identical to zero have been collapsed into a single one for readability.}
    \label{fig:cumulene-cutoff}
\end{figure}

In \cref{fig:cumulene-cutoff}, we present the curves related to \cref{tab:cumulene-cutoffs}. All models where the dihedral information is accessible can resolve the benchmark, but vary in match to the ground truth. In particular, we note that models with two short-range message passing steps perform better than a single one. This may be due to the high degree of degeneracy in atomic environments at short cutoffs, which is alleviated by message passing.

\clearpage
\section{Hyper-parameter sweep for the \ch{NaCl} cluster and \ch{MgO} surface datasets}
\label{sec:apx-valset}

We report the results of a hyper-parameter sweep for  the \ch{NaCl} cluster and \ch{MgO} surface datasets on a \scare{inner} train/validation split of the training set used for the experiments in \cref{sec:experiments}, aiming to gauge the impact of hyper-parameter tuning and early stopping on the validation set used for \cref{tab:metrics}.

We used random train/validation split ($4000$ samples / $500$ samples) of the \ch{NaCl} cluster and \ch{MgO} surface training datasets, sweeping the following combinations of parameters: Initial learning rate $(0.001, 0.0001)$, learning rate decay (linear, exponential), batch size $(32, 64)$, and energy/force loss weights $(0.5/0.5, 100/1, 1000/1)$ over $4000$ epochs of training with the ADAM optimizer.
From the resulting models, we selected those with the best \scare{inner} validation RMSE on energy and forces and evaluated the models on the \scare{outer} validation split used for \cref{tab:metrics}.
The results  are presented in \cref{tab:hp_tuning_mgo} and \cref{tab:hp_tuning_nacl}.
The model selected by the best energy error on the validation set performs best in both cases, similar to the model used in the main text. The model with the best force error ranks second for the \ch{MgO} surface task. We also verified that both models are able to reproduce the curves in \cref{fig:aumgo+nacl}.
In summary, explicitly tuning the model on an inner train/validation split, including performing early stopping on this validation set, does not significantly impact benchmark results as reported in \cref{tab:metrics}.

\begin{table}[h]
    \caption{Errors on the validation set for the hyperparameter set with the best inner validation metrics for the \ch{MgO} task.}
    \centering
    \begin{tabular}{r|cc}
\toprule
Model & Test RMSE $E$ (meV/atom) & Test RMSE $\F$ (meV/\unit{\angstrom}) \\
\midrule
\lorem & 0.064 & 4.076 \\
Nearest other    & 0.071 (\cace) & 5.971 (\mace) \\
Best $E$ model   & 0.064 & 5.229 \\
Best $F$ model   & 0.078 & 3.630 \\
\bottomrule
\end{tabular}

    \label{tab:hp_tuning_mgo}
\end{table}

\begin{table}[h]
    \caption{Errors on the validation set for the hyperparameter set with the best inner validation metrics for the \ch{NaCl} task.}
    \centering
    \begin{tabular}{r|cc}
\toprule
Model & Test RMSE $E$ (meV/atom) & Test RMSE $\F$ (meV/\unit{\angstrom}) \\
\midrule
\lorem  & 0.112 & 1.155 \\
Nearest other  & 0.210 (\cace) & 9.784 (\cace) \\
Best $E$ model & 0.076 & 2.613 \\
Best $F$ model & 0.101 & 1.473 \\
\bottomrule
\end{tabular}

    \label{tab:hp_tuning_nacl}
\end{table}

\clearpage
\CL{
\section{Result variants: MAE metrics, RMSE biodimers forces, different seeds}
\label{sec:apx-mae-rmse}

\Cref{tab:metrics} in the main text reports root mean squared errors (RMSE). For completeness, we provide the corresponding mean absolute errors (MAE) in \cref{tab:metrics-mae}. Additionally, \cref{fig:bio-dimers-rmse} shows the biodimers force errors using RMSE (the main text uses MAE). The conclusions are unchanged: \lorem achieves the lowest or competitive errors across all datasets and dimer classes.

\CL{To assess sensitivity to random initialization and dataset shuffling, we also report results for an alternative seed for \lorem ($\dagger$) in \cref{tab:metrics-seed} (RMSE) and \cref{tab:metrics-mae-seed} (MAE). This represents the worse of two seeds trained per task. The results are consistent with the main table, confirming that \lorem's performance is robust to the choice of random seed. We also confirmed that qualitative performance on the benchmark tasks is similar.}

\begin{table}[h]
    \caption{\CL{Mean absolute errors for energy $E$ and forces $\F$. Models for which MAE metrics are not available (4G-NN, \textsc{SpookyNet}) are omitted. Best in bold, second best underlined. See \cref{tab:metrics} for details.}}
    \centering
    \begin{tabular}{r | r r r r}
\toprule
                 Dataset  &  \makecell{\textsc{Lorem} \\{ $1\times$SR$+$LR}}  &  \makecell{\textsc{Cace-Les} \\{ $1\times$SR$+$LR}}  &  \makecell{\textsc{Mace} \\{ $2\times$SR}}  &  \makecell{\textsc{Pet} \\{ $2\times$ SR}} \\ 
\midrule
\ch{MgO} surface\hspace{3mm} $E$ (meV/at)  &   \underline{\num{0.041}}  &      \textbf{\num{0.036}}  &               \num{0.353}  &               \num{0.188} \\ 
{\small (Validation set)}\hspace{3mm} $\F$ (meV/Å)  &      \textbf{\num{2.869}}  &               \num{5.584}  &   \underline{\num{3.620}}  &               \num{4.191} \\ 
\midrule
\ch{NaCl} cluster\hspace{3mm} $E$ (meV/at)  &      \textbf{\num{0.090}}  &   \underline{\num{0.161}}  &               \num{1.345}  &               \num{1.249} \\ 
{\small (Validation set)}\hspace{3mm} $\F$ (meV/Å)  &      \textbf{\num{0.778}}  &   \underline{\num{6.388}}  &              \num{23.598}  &              \num{20.011} \\ 
\midrule
Biodimers\hspace{3mm} $E$ (meV/at)  &      \textbf{\num{0.155}}  &   \underline{\num{0.566}}  &               \num{3.220}  &               \num{2.657} \\ 
\hspace{3mm} $\F$ (meV/Å)  &      \textbf{\num{1.023}}  &   \underline{\num{1.787}}  &               \num{5.278}  &               \num{5.352} \\ 
\midrule
Cumulene\hspace{3mm} $E$ (meV/at)  &      \textbf{\num{1.096}}  &              \num{14.567}  &               \num{8.961}  &   \underline{\num{1.363}} \\ 
\hspace{3mm} $\F$ (meV/Å)  &  \underline{\num{22.078}}  &             \num{107.946}  &              \num{70.467}  &     \textbf{\num{13.610}} \\ 
\bottomrule
\end{tabular}

    \label{tab:metrics-mae}
\end{table}

\begin{table}[h]
    \caption{\CL{RMSE for energy $E$ and forces $\F$, with \lorem results for an alternative seed ($\dagger$), representing the worse of two seeds. Best in bold, second best underlined. See \cref{tab:metrics} for details.}}
    \centering
    \begin{tabular}{r | r r r r r r}
\toprule
                 Dataset  &  \makecell{\textsc{Lorem}$^{\dagger}$ \\{ $1\times$SR$+$LR}}  &  \makecell{\textsc{Cace-Les} \\{ $1\times$SR$+$LR}}  &  \makecell{\textsc{Mace} \\{ $2\times$SR}}  &  \makecell{\textsc{Pet} \\{ $2\times$ SR}}  &  \makecell{4G-NN \\{ $1\times$SR$+$LR}}  &  \makecell{\textsc{SpookyNet} \\{ $6\times$SR$+$LR}} \\ 
\midrule
\ch{MgO} surface\hspace{3mm} $E$ (meV/at)  &      \textbf{\num{0.065}}  &   \underline{\num{0.071}}  &               \num{0.376}  &               \num{0.210}  &               \num{0.219}  &               \num{0.107} \\ 
{\small (Validation)}\hspace{3mm} $\F$ (meV/Å)  &      \textbf{\num{4.381}}  &               \num{7.913}  &               \num{5.971}  &               \num{6.261}  &              \num{66.000}  &   \underline{\num{5.337}} \\ 
\midrule
\ch{NaCl} cluster\hspace{3mm} $E$ (meV/at)  &      \textbf{\num{0.112}}  &               \num{0.210}  &               \num{1.681}  &               \num{1.517}  &               \num{0.481}  &   \underline{\num{0.135}} \\ 
{\small (Validation)}\hspace{3mm} $\F$ (meV/Å)  &   \underline{\num{1.275}}  &               \num{9.784}  &              \num{40.219}  &              \num{42.438}  &              \num{32.780}  &      \textbf{\num{1.052}} \\ 
\midrule
Biodimers\hspace{3mm} $E$ (meV/at)  &      \textbf{\num{0.370}}  &   \underline{\num{2.259}}  &               \num{7.793}  &               \num{6.758}  &                        --  &                        -- \\ 
\hspace{3mm} $\F$ (meV/Å)  &      \textbf{\num{1.985}}  &   \underline{\num{3.163}}  &              \num{16.150}  &              \num{16.470}  &                        --  &                        -- \\ 
\midrule
Cumulene\hspace{3mm} $E$ (meV/at)  &   \underline{\num{3.307}}  &              \num{17.803}  &              \num{12.592}  &      \textbf{\num{3.205}}  &                        --  &                        -- \\ 
\hspace{3mm} $\F$ (meV/Å)  &  \underline{\num{55.412}}  &             \num{147.616}  &             \num{104.318}  &     \textbf{\num{46.905}}  &                        --  &                        -- \\ 
\bottomrule
\end{tabular}

    \label{tab:metrics-seed}
\end{table}

\begin{table}[h]
    \caption{\CL{MAE for energy $E$ and forces $\F$, with \lorem results for an alternative seed ($\dagger$), representing the worse of two seeds. Models for which MAE metrics are not available (4G-NN, \textsc{SpookyNet}) are omitted. Best in bold, second best underlined. See \cref{tab:metrics} for details.}}
    \centering
    \begin{tabular}{r | r r r r}
\toprule
                 Dataset  &  \makecell{\textsc{Lorem}$^{\dagger}$ \\{ $1\times$SR$+$LR}}  &  \makecell{\textsc{Cace-Les} \\{ $1\times$SR$+$LR}}  &  \makecell{\textsc{Mace} \\{ $2\times$SR}}  &  \makecell{\textsc{Pet} \\{ $2\times$ SR}} \\ 
\midrule
\ch{MgO} surface\hspace{3mm} $E$ (meV/at)  &   \underline{\num{0.040}}  &      \textbf{\num{0.036}}  &               \num{0.353}  &               \num{0.188} \\ 
{\small (Validation set)}\hspace{3mm} $\F$ (meV/Å)  &      \textbf{\num{3.133}}  &               \num{5.584}  &   \underline{\num{3.620}}  &               \num{4.191} \\ 
\midrule
\ch{NaCl} cluster\hspace{3mm} $E$ (meV/at)  &      \textbf{\num{0.090}}  &   \underline{\num{0.161}}  &               \num{1.345}  &               \num{1.249} \\ 
{\small (Validation set)}\hspace{3mm} $\F$ (meV/Å)  &      \textbf{\num{0.882}}  &   \underline{\num{6.388}}  &              \num{23.598}  &              \num{20.011} \\ 
\midrule
Biodimers\hspace{3mm} $E$ (meV/at)  &      \textbf{\num{0.322}}  &   \underline{\num{0.566}}  &               \num{3.220}  &               \num{2.657} \\ 
\hspace{3mm} $\F$ (meV/Å)  &      \textbf{\num{1.110}}  &   \underline{\num{1.787}}  &               \num{5.278}  &               \num{5.352} \\ 
\midrule
Cumulene\hspace{3mm} $E$ (meV/at)  &      \textbf{\num{1.212}}  &              \num{14.567}  &               \num{8.961}  &   \underline{\num{1.363}} \\ 
\hspace{3mm} $\F$ (meV/Å)  &  \underline{\num{28.448}}  &             \num{107.946}  &              \num{70.467}  &     \textbf{\num{13.610}} \\ 
\bottomrule
\end{tabular}

    \label{tab:metrics-mae-seed}
\end{table}

\begin{figure}[h]
    \centering
    \includegraphics[scale=0.65]{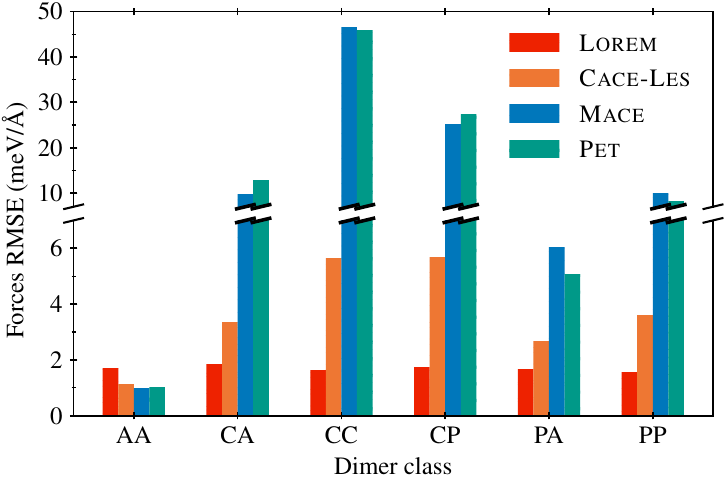}
    \caption{Root mean squared error on forces for different models on the different dimer classes. Compare with \cref{fig:bio_dimers} in the main text, which shows MAE.}
    \label{fig:bio-dimers-rmse}
\end{figure}
}

\clearpage
\CL{
\section{ADAPT benchmark}
\label{sec:apx-adapt}

To evaluate \lorem on a larger-scale dataset, we trained it on the ADAPT silicon point-defect dataset \citep{dxzhrjk2025pre}. This dataset contains \num{206973} training, \num{51743} validation, and \num{4425} test structures, each consisting of \num{217} atoms (\num{215} Si atoms plus \num{2} dopant atoms drawn from \num{24} elements, forming \num{52} unique dopant pairs). The test set consists of the first frames from \num{100} held-out relaxation trajectories, following the evaluation protocol of the original work.

We used \lorem with default model hyperparameters ($\cutoff=\qty{5}{\angstrom}$, $\lmax=6$, $\lrlmax=2$, \num{128} scalar features, \num{8} spherical channels, \num{32} radial basis functions).
Training hyperparameters were selected via a sweep of \num{24} combinations on a \num{9500}-structure debug subset (\num{100} epochs): optimizer (Muon \cite{jordan2024muon}, ADAM), initial learning rate $(\num{1e-3}, \num{1e-4})$, learning rate decay (linear, exponential), and energy/force loss weights $(0.5/0.5, 100/1, 1000/1)$.
Four configurations from the Pareto frontier were selected for production training on the full dataset, all using the Muon optimizer with initial learning rate \num{1e-3}: energy/force loss weights $0.5/0.5$ with linear and exponential decay, and $100/1$ with linear and exponential decay.
These were trained for \num{300} epochs (\qty{65}{h} on a single H100 GPU).

Results are shown in \cref{tab:adapt}, reporting the best force MAE and best energy MAE across the four production runs, each evaluated at the checkpoint with the highest summed $R^2$ of energy and forces on the validation set.
\lorem nearly matches the force accuracy of the ADAPT model (which uses an all-to-all transformer architecture with $\sim$\qty{4}{M} parameters) while achieving substantially better energy accuracy---despite using a single model for both properties, compared to ADAPT's separate energy-only model. \lorem also significantly outperforms all other baselines, including MACE (both retrained and foundation models) and MatterSim.

\begin{table}[h]
    \caption{\CL{Mean absolute errors on the \num{100}-structure ADAPT benchmark test set. Baseline results are from \citet{dxzhrjk2025pre}. The ADAPT model uses a separate energy-only model for energies; all other models predict both properties jointly. Best in bold, second best underlined. Parameter counts: ADAPT Small $\sim$\qty{4}{M}, ADAPT Large $\sim$\qty{18}{M}.}}
    \centering
    \begin{tabular}{r | r r}
    \toprule
    Model & $\F$ MAE (\unit{eV\per\angstrom}) & $E$ MAE (\unit{eV}) \\
    \midrule
    \lorem (best $\F$) & \underline{\num{0.0128}} & \underline{\num{0.097}} \\
    \lorem (best $E$)  & \num{0.0136} & \textbf{\num{0.076}} \\
    \midrule
    ADAPT Small        & \textbf{\num{0.0126}} & \num{0.578} \\
    ADAPT Large        & \num{0.0136} & -- \\
    \mace Retrained    & \num{0.0217} & \num{1.313} \\
    \mace MP0a Large   & \num{0.0439} & \num{6.101} \\
    \mace MPA-0 Medium & \num{0.0349} & \num{2.048} \\
    \mace OMAT-0 Medium & \num{0.0283} & \num{3.223} \\
    MatterSim 1M       & \num{0.0323} & \num{1.743} \\
    MatterSim 5M       & \num{0.0335} & \num{0.829} \\
    \bottomrule
    \end{tabular}
    \label{tab:adapt}
\end{table}
}

\end{document}